\documentclass[%
aip,
amsmath,amssymb,
reprint,%
]{revtex4-1}

\usepackage{graphicx}
\usepackage{dcolumn}
\usepackage{bm}

\usepackage[utf8]{inputenc}
\usepackage[T1]{fontenc}
\usepackage{mathptmx}
\usepackage{etoolbox}

\usepackage{amsmath}
\makeatletter
\def\@email#1#2{%
	\endgroup
	\patchcmd{\titleblock@produce}
	{\frontmatter@RRAPformat}
	{\frontmatter@RRAPformat{\produce@RRAP{*#1\href{mailto:#2}{#2}}}\frontmatter@RRAPformat}
	{}{}
}%
\makeatother

\begin{document}
	
\preprint{AIP/123-QED}

\title[]{Novel nonlinear system family generated from coupling effect of Sin-Cosine function}

\author{Fangfang Zhang}
\email{Fangfang Zhang: zhff4u@qlu.edu.ch}
\affiliation{School of Information and Automation Engineering, Qilu University of Technology (Shandong Academy of Sciences), Jinan 250353, PR China
}

\author{Jinyi Ge}
\email{Jinyi Ge: 10431230550@stu.qlu.edu.cn}
\affiliation{School of Information and Automation Engineering, Qilu University of Technology (Shandong Academy of Sciences), Jinan 250353, PR China
}

\author{Cuimei Jiang*}
\email{Corresponding author. Cuimei Jiang: jcm2017@qlu.edu.cn}
\affiliation{ School of Mathematics and Statistics, Qilu University of Technology (Shandong Academy of Sciences), Jinan 250353, PR China 
}

\author{Han Bao}
\email{Han Bao: hanbao@cczu.edu.cn}
\affiliation{School of Microelectronics and Control Engineering, Changzhou University, Changzhou 213159, PR China 
}

\author{Jianlin Zhang}
\email{Jianlin Zhang: s22060809029@smail.cczu.edu.cn}
\affiliation{School of Microelectronics and Control Engineering, Changzhou University, Changzhou 213159, PR China 
}

\author{Da Wang}
\email{Da Wang: wangda@sdnu.edu.cn}
\affiliation{Business School, Research Center of Dynamics System and Control Science, Shandong Normal University, Jinan 250014, PR China
}

\author{Yang Zhao}
\email{Yang Zhao: zdh1136@126.com}
\affiliation{School of Information and Automation Engineering, Qilu University of Technology (Shandong Academy of Sciences), Jinan 250353, PR China
}

\date{\today}

\begin{abstract}
Based on the coupling effect of the Sine-Cosine function, we discover a novel class of nonlinear system family, which is called Sine-Cosine Nonlinear System Family (SCNSF). This system family exhibits chaotic characteristics in the real number domain and fractal characteristics in the complex number domain. Firstly, we propose the specific classification of SCNSF and provide their general mathematical description. Secondly, we propose three types of classic systems of SCNSF and focus on their chaotic characteristics in the real number domain and the hardware implementation. The simulation experiments and physical experiments show that these three types of systems all exhibit a large chaotic range and posses broad application prospects. Then, the mechanism of generating chaos based on the coupling effect of Sine-Cosine function is discovered. Finally, it is found that the above-mentioned three types of systems exhibit fractal characteristics in the complex number domain with high parameter sensitivity. Simulation experiments are carried out, and the results show that SCNSF demonstrates a wide range of fractal characteristics and the results visually display the Julia Set and the distribution of periodic points, demonstrating that the SCNSF possesses a wide range of fractal characteristics. The chaotic and fractal properties of SCNSF offer significant potential for further applications in fields such as chaotic circuit design, information encryption and signal detection.
\end{abstract}

\maketitle
\begin{quotation}
The Sine-Cosine function, which is widely adopted in mathematics and physics, has attracted our attention due to its unique properties. By delving into the coupling effect of the Sine-Cosine function, we discover a previously unreported class of nonlinear systems, namely the Sine-Cosine Nonlinear System Family (SCNSF). This discovery is motivated by the need to expand the repertoire of nonlinear systems and understand the complex behaviors that can emerge from the combination of basic trigonometric functions. The SCNSF has both chaotic characteristics in the real number domain and fractal characteristics in the complex number domain. The classification and general mathematical description of SCNSF provide a solid theoretical foundation for further research. The proposal of three types of classic systems within SCNSF and the investigation of their chaotic properties and hardware implementation open up new avenues for practical applications. The large chaotic range exhibited by these systems implies their potential applications in various fields such as secure communication and chaotic circuit design. Moreover, the discovery of the chaos generation mechanism based on the coupling effect of the Sine-Cosine function deepens our understanding of the origin of chaos. In the complex number domain, the high parameter sensitivity and rich fractal patterns of SCNSF can be can be harnessed to develop more advanced encryption algorithms and more sensitive signal detection methods, thereby contributing to the advancement of information security and signal processing technologies. Overall, the chaotic and fractal properties of SCNSF make it a valuable asset in the pursuit of innovative solutions in multiple scientific and engineering disciplines.
\end{quotation}

\section{\label{sec:level1}Introduction}

Despite being governed by deterministic rules, chaotic systems have significant advantages and broad application prospects in scientific research and engineering. Their unique dynamic behaviors, such as randomness and long-term unpredictability, make them well-suited for high-quality random number generation, encryption, and secure communications. Their ergodicity allows chaotic systems to cover a wide range of states within a short time, facilitating global optimization. The sensitivity to initial conditions amplifies small changes in the system, which in turn enhances its capabilities in pattern recognition and fault diagnosis. As a result, chaotic systems have become a major focus of current research.

At present, the research on chaotic systems mainly includes three directions: (1) The proposal of new chaotic systems and their dynamical analysis; (2) The synchronization and control of chaotic systems; (3) The applications of chaotic systems such as encryption.

As for the construction of new chaotic systems, researchers continuously explore various models and methods to develop chaotic systems with specific properties\cite{OffsetBoosting}. For instance, Zhu et al\cite{ZHU2023113370} proposed a novel closed-loop multiple modulation coupling (CMMC) method to construct hyperchaotic systems with three positive Lyapunov exponents, designed a grid multi-cavity model, and implemented pseudorandom number generators for hardware applications with high complexity and rich dynamics. Bao et al.\cite{Bao2010TransientCI,Bao2011ChaoticMC} designed a smooth memristor oscillator and studied the transition from transient chaos to stable chaos and intermittent periodic dynamics.  Xu et al.\cite{Xu2021ContinuousNM} proposed a Rulkov chaotic model with different firing patterns and stochastic resonance, observed its dynamic response and developed a hardware implementation scheme based on analog circuits. Bao et al.\cite{Bao2023SineTransformBasedMH} proposed a novel two-dimensional sine transform memristor model, and designed six types of pseudorandom number generators. To address the common issues of frail chaos and dynamical degradation, Zhu et al\cite{Liu10423990}. proposed a delayed feedback method to construct hyperchaotic maps, enhancing complexity, dimensionality, and period length, and applied it to pseudorandom number generation and image compression on FPGA platforms. 

As for the study of synchronization and control of chaotic systems, many researchers aim to understand the interactions between chaotic systems in complex networks and develop control strategies to achieve synchronization, including complete synchronization (CS)\cite{MahmoudNonlinea, PARK2005579, ZHU2024115281}, projective synchronization (PS) and modified projective synchronization (MPS)\cite{MAHMOUD20102286}, anti-synchronization (AS)\cite{LIU20113046}, lag synchronization (LS)\cite{Lagsynchronizat}, modified projective phase synchronization (MPPS)\cite{MAHMOUD201369}, complex function projective synchronization (CFPS)\cite{plexfunction}, combination synchronization\cite{SunCombination}, complex modified function projective synchronization\cite{LIU2017440, LIANG2024129516}, etc.

In the application of chaotic systems, researchers utilize these systems in various fields such as cryptography\cite{ZHOU2023782,9431651,Ma2024ASM,Huang2023ARI}, secure communications\cite{Zhang2021}, and optimization\cite{Bao2024InitialOffsetControlCH,Zhang2023ChaoticNN}. Mengxin Jin\cite{JinHyperchaos} introduced hyperchaotic systems and their importance in cryptography, communication and control, as well as research advancements in the generation, synchronization, and extreme multistability of complex chaotic systems.

Clearly, the chaotic systems proposed in the above literatures are numerous and diverse, and have not been systematically classified. At the same time, the physical implementation of these systems is relatively complex. Additionally, most literatures have discussed the chaotic characteristics of these systems, but have not considered the fractal characteristics exhibited by the chaotic systems in the complex field. 

Fractals, as an important geometric manifestation, are exhibited for many chaotic systems in the complex field , including self-similarity and infinite complexity. Fractals, through self-similarity, reveal the structural laws of complex natural phenomena, aiding in the more precise description and prediction of the behavior of complex systems\cite{WANG2023114175,singh2024,luo2024}. Fractal structures not only have signiﬁcant theoretical value in mathematics but also demonstrate broad potential in practical applications. For instance, fractal geometry has signiﬁcant applications in ﬁelds such as image processing\cite{LI2023212236}, signal compression\cite{JANGBAHADURSAINI2023100698}, and texture analysis\cite{JORDANOVA2022100108}.

In 2023, we discovered that the coupling effect of Sine and Cosine function can induce chaotic phenomena, leading us to propose a two-dimensional sine-cosine chaotic system and apply it to encryption\cite{10481317}. When these functions are coupled with each other or combined with existing chaotic systems, their inherent periodicity is disrupted, exhibiting chaotic in the real number domain and fractal characteristics in the complex number domain. Therefore, in this article, we propose a collection of various nonlinear systems with such a mechanism, which is defined as the Sine-Cosine Nonlinear System Family (SCNSF). The contributions and main work of this paper are as follows:

(1) Based on the coupling effect of sine and cosine functions, we propose the definition and general mathematical description of SCNSF, which includes several types of sine-cosine nonlinear systems. These systems are extremely simple but possess extremely rich dynamic characteristics and are easy for physical implementation. 

(2) Three novel classes of sine-cosine nonlinear systems are proposed, including the Sine-Cosine Discrete System (SCDS), the Multidimensional Chebyshev System (MDCS), and the Sine-Logistic System (SLS). The chaotic characteristics of these three systems in the real domain are studied, sufficient conditions for the stability of fixed points are provided, hardware implementations are carried out, and the mechanisms by which these systems generate chaos are proposed.

(3) The fractal characteristics and parameter sensitivity of these three types of sine-cosine nonlinear systems in the complex domain are studied, and the mechanisms by which these systems generate fractals are analyzed.

The structure of the paper is as follows: Section 2 proposes the SCNSF. Section 3 presents the three types of sine-cosine nonlinear systems, analyzes their chaotic characteristics and stability in the real domain, and discusses their hardware implementation, while also proposing the mechanisms of generating chaos. In Section 4, we explore the fractal characteristics of these three types of sine-cosine nonlinear systems in the complex domain and analyzes the mechanisms behind their fractal generation. Section 5 concludes the paper.

\section{\label{sec:level1}Sine-Cosine Nonlinear System Family}

Sine and cosine functions exhibit periodicity and they are easily realizable in physical systems. When they are coupled with each other or combined with existing chaotic systems, their inherent periodicity is disrupted, displaying chaotic and fractal characteristics. This novel system family is referred to as the "Sine-Cosine Nonlinear System Family" (SCNSF).

Under the condition of satisfying the composite function, the general mathematical description of the $m$-dimensional SCNSF is as follows,
\begin{eqnarray}
\left\{ \begin{array}{l}
	x_{1,(n + 1)} = {\alpha_1}g\big({\beta_1}f(x_{2,(n)})+{\gamma_1}\big),\\
	x_{2,(n + 1)} = {\alpha_2}g\big({\beta_2}f(x_{3,(n)})+{\gamma_2}\big),\\
	......\\
	x_{{m - 1},(n + 1)} = {\alpha_{m - 1}}g\big({\beta_{m - 1}}f(x_{m,(n)})+{\gamma_{m - 1}}\big),\\
	x_{m,(n + 1)} = {\alpha_m}g\big({\beta_m}f(x_{1,(n)})+{\gamma_m}\big),
\end{array} \right.
\label{eq:sccs}
\end{eqnarray}
where $\alpha_1, \alpha_2, \ldots, \alpha_m$, $\beta_1, \beta_2, \ldots, \beta_m$ and $\gamma_1, \gamma_2, \ldots, \gamma_m$ are parameters, and $x_{i,(n)}$ denote the value of the $i$-th state variable at the $n$-th iteration, and $x_{i,(n+1)}$ denote the value of the $i$-th state variable at the $(n+1)$-th iteration. $f$ and $g$ represent functions or mappings  where one is a sine-cosine function and the other is an ordinary linear or nonlinear function.

In the following text, we will separately investigate their chaotic characteristics in the real number domain and the fractal features in the complex number domain.

\section{\label{sec:level1}Sine-Cosine Nonlinear System Family in real Field}

We combine some types of mappings and classic chaotic systems with sine and cosine functions, and propose three types of new sine-cosine nonlinear systems.

\subsection{\label{sec:level2}Sine-Cosine Discrete System in real Field}
The mathematical description of a $m$-dimensional Sine-Cosine System is represented as
\begin{eqnarray}
\left\{ \begin{array}{l}
	x_{1,(n + 1)} = {\alpha_1}g\big(x_{2,(n)}\big),\\
	x_{2,(n + 1)} = {\alpha_2}g\big(x_{3,(n)}\big),\\
	......\\
	x_{{m - 1},(n + 1)} = {\alpha_{m - 1}}g\big(x_{m,(n)}\big),\\
	x_{m,(n + 1)} = {\alpha_m}g\big(x_{1,(n)}\big),
\end{array} \right.\
\label{eq:scs}
\label{scs}
\end{eqnarray}
where $g$ is a sine or cosine function. It is a speical case of SCNSF (\ref{eq:sccs}) with $\beta_1=\beta_2= \ldots=\beta_m=1$, $\gamma_1=\gamma_2=\ldots= \gamma_m=0$ and $f(x)=x$.
Here we explore the dynamical characteristics of a three-dimensional system as a case study. By setting $g(x) = \sin(x)$, the three-dimensional Sine-Cosine Discrete System (3D-SCDS) is represented by the following equation:
\begin{eqnarray}
\left\{ \begin{array}{l}
	{x_{(n + 1)}} = a\sin ({y_{(n)}}),\\
	{y_{(n + 1)}} = b\sin ({z_{(n)}}),\\
	{z_{(n + 1)}} = c\sin ({x_{(n)}}).
\end{array} \right.\
\label{eq:scs2}
\end{eqnarray}
where $a,b,c$ are parameters and their values are positive.
\begin{figure}
\centering
\includegraphics[width=0.9\linewidth]{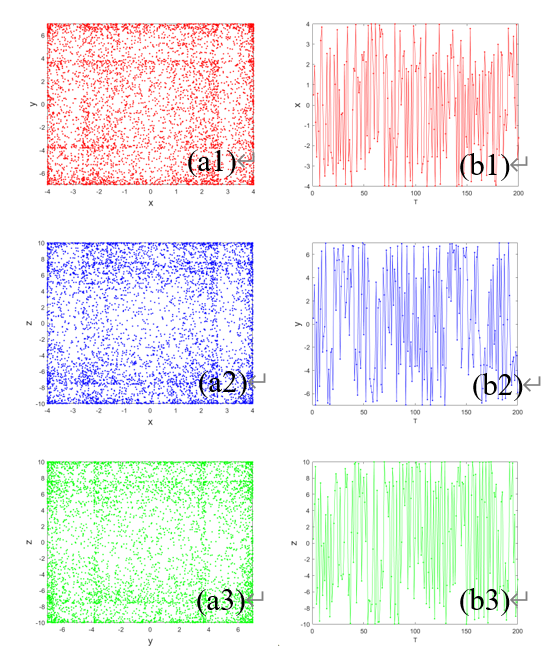}
\caption{The phase portrait and time-domain diagram of 3D-SCDS (\ref{eq:scs2}). (a1) $x-y$ (a2) $x-z$ (a3) $y-z$  (b1) $t-x$ (b2) $t-y$ (b3) $t-z$ .}
\label{fig:sx1}
\end{figure}
When $a=4$, $b=7$, and $c=10$,  the phase portrait and time-domain diagram of the 3D-SCDS (\ref{eq:scs2}) can be depicted in Fig.\ref{fig:sx1}, where Fig.\ref{fig:sx1}(a1), Fig.\ref{fig:sx1}(a2) and Fig.\ref{fig:sx1}(a3) represent the phase portrait of $x,y,z$, and Fig. \ref{fig:sx1}(b1), Fig.\ref{fig:sx1}(b2) and Fig.\ref{fig:sx1}(b3) represent  time-domain diagram of $x,y,z$. 
From Fig.\ref{fig:sx1}, it can be observed that the output of the 3D-SCDS neither converges nor diverges, and it lacks periodicity. Therefore, it is preliminarily inferred that the 3D-SCDS (3) is chaotic. 

\begin{figure}
\centering
\includegraphics[width=0.8\linewidth]{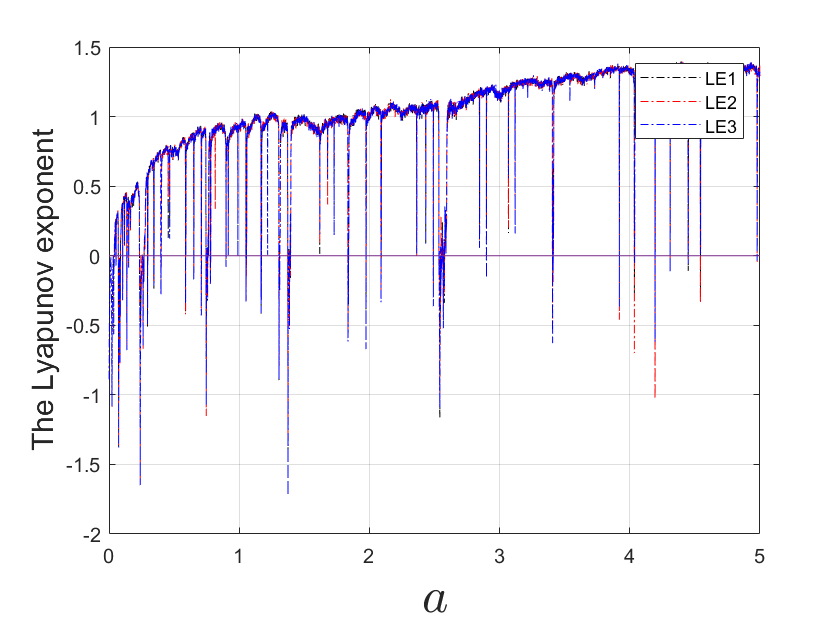}
\caption{Lyapunov exponents of 3D-SCDS(\ref{eq:scs2})}
\label{fig:SCS Lyapunov exponent plot}
\end{figure}

\begin{figure}
\centering
\includegraphics[width=0.8\linewidth]{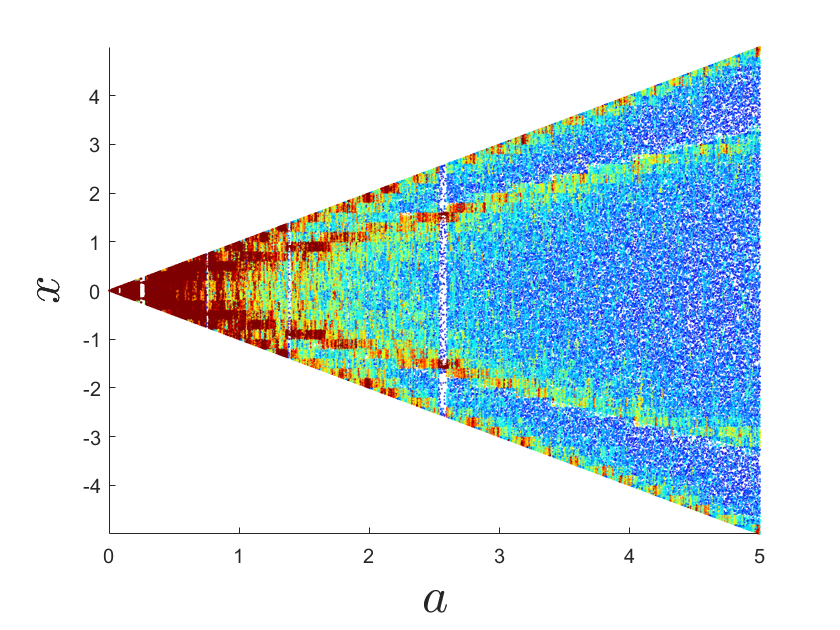}
\caption[]{The bifurcation diagram of 3D-SCDS(\ref{eq:scs2})}
\label{fig:SCS BifurcationGragh}
\end{figure}

The Lyapunov exponents (LEs) are further given in Fig.\ref{fig:SCS Lyapunov exponent plot}. The bifurcation diagram of system is depicted in Fig.\ref{fig:SCS BifurcationGragh}, where $b=7$, $c=10$ and $0\le a\le 5$. From Fig.\ref{fig:SCS Lyapunov exponent plot} and Fig.\ref{fig:SCS BifurcationGragh}, there are positive LE. The chaotic range of 3D-SCDS gradually expands as the parameter $a$ increases, which leads to the conclusion that the system is chaotic.  

Next, we will compute the fixed points of Eq.(\ref{eq:scsfixed}), i.e,
\begin{eqnarray}
\left\{ \begin{array}{l}
	x_f = a\sin y_f,\\
	y_f = b\sin z_f,\\
	z_f = c\sin x_f.
\end{array} \right.
\label{eq:scsfixed}
\end{eqnarray}
Where ($x_f$, $y_f$, $z_f$) denotes the fixed point. Simplify it to an equation involving only $x_f$ as follows,
\begin{eqnarray}
	x_f = a \sin \left( b \sin \left( c \sin x_f \right) \right).
\end{eqnarray}
Then, the fixed points of the system can be obtained by solving
\begin{eqnarray}
h(x_f) = a\sin ( b\sin (c\sin x_f)) - x_f =0.
\end{eqnarray}

\begin{figure}
\centering
\includegraphics[width=0.8\linewidth]{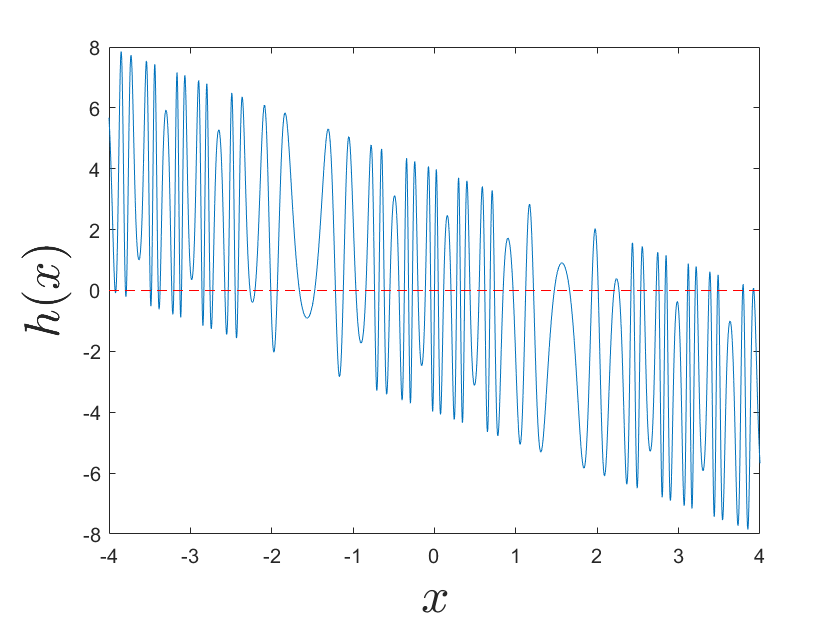}
\caption[]{Fixed points of 3D-SCDS(\ref{eq:scs2})}
\label{fig:Fixed Points of SCS}
\end{figure}

The graph of function $h(x_f)$ is displayed in Fig.\ref{fig:Fixed Points of SCS}, where the intersection points between the blue line indicating $h(x_f)$ and the red line $x_f=0$ represent the fixed points of 3D-SCDS (\ref{eq:scs2}). It can be observed that the system has multiple fixed points when $a=4$, $b=7$ and $c=10$. Next, we analyze the stability of these fixed points.

Firstly, we calculate the Jacobian matrix of Eq.(\ref{eq:scs2}) as follows.

\begin{eqnarray}
\begin{array}{l}
	J = {\left. {\left[ {\begin{array}{*{20}{c}}
					{\frac{{\partial {f_1}}}{{\partial x}}}&{\frac{{\partial {f_1}}}{{\partial y}}}&{\frac{{\partial {f_1}}}{{\partial z}}}\\
					{\frac{{\partial {f_2}}}{{\partial x}}}&{\frac{{\partial {f_2}}}{{\partial y}}}&{\frac{{\partial {f_2}}}{{\partial z}}}\\
					{\frac{{\partial {f_3}}}{{\partial x}}}&{\frac{{\partial {f_3}}}{{\partial y}}}&{\frac{{\partial {f_3}}}{{\partial z}}}
			\end{array}} \right]} \right|_{({x_f},{y_f},{z_f})}}\\
	= {\left. {\left[ {\begin{array}{*{20}{c}}
					0&{a\cos y}&0\\
					0&0&{b\cos z}\\
					{c\cos x}&0&0
			\end{array}} \right]} \right|_{({x_f},{y_f},{z_f})}}
\end{array}
\label{eq:SCSJ}
\end{eqnarray}
			
Then, we solve for the eigenvalues $\lambda_1$ , $\lambda_2$ and $\lambda_3$ of the Jacobian matrix, and get 
\begin{eqnarray}
\left| {J - \lambda I} \right| = abc\cos x_f\cos y_f\cos z_f - {\lambda ^3} = 0.
\label{eq:SCSJ2}
\end{eqnarray}
It can be concluded that
\begin{eqnarray}
{\lambda ^3} = abc\cos x_f\cos y_f\cos z_f,
\label{eq:SCSJ3}
\end{eqnarray}

Finally, according to the stability theorem of discrete systems, it can be known that when
\begin{eqnarray}
\left| \lambda  \right| = \left| \sqrt[3]{{abc\cos x_f\cos y_f\cos z_f}} \right| < \left| \sqrt[3]{{abc}} \right| < 1,
\label{eq:SCSJ4}
\end{eqnarray}
this fixed point is stable. Therefore, we obtain the following sufﬁcient conditions for
the stability of fixed points.

\textbf{Theorem 1:} For the 3D-SCDS (\ref{eq:scs2}), if $\left| abc \right| < 1$, so these have stable fixed points.

\begin{figure}
\centering
\includegraphics[width=1\linewidth]{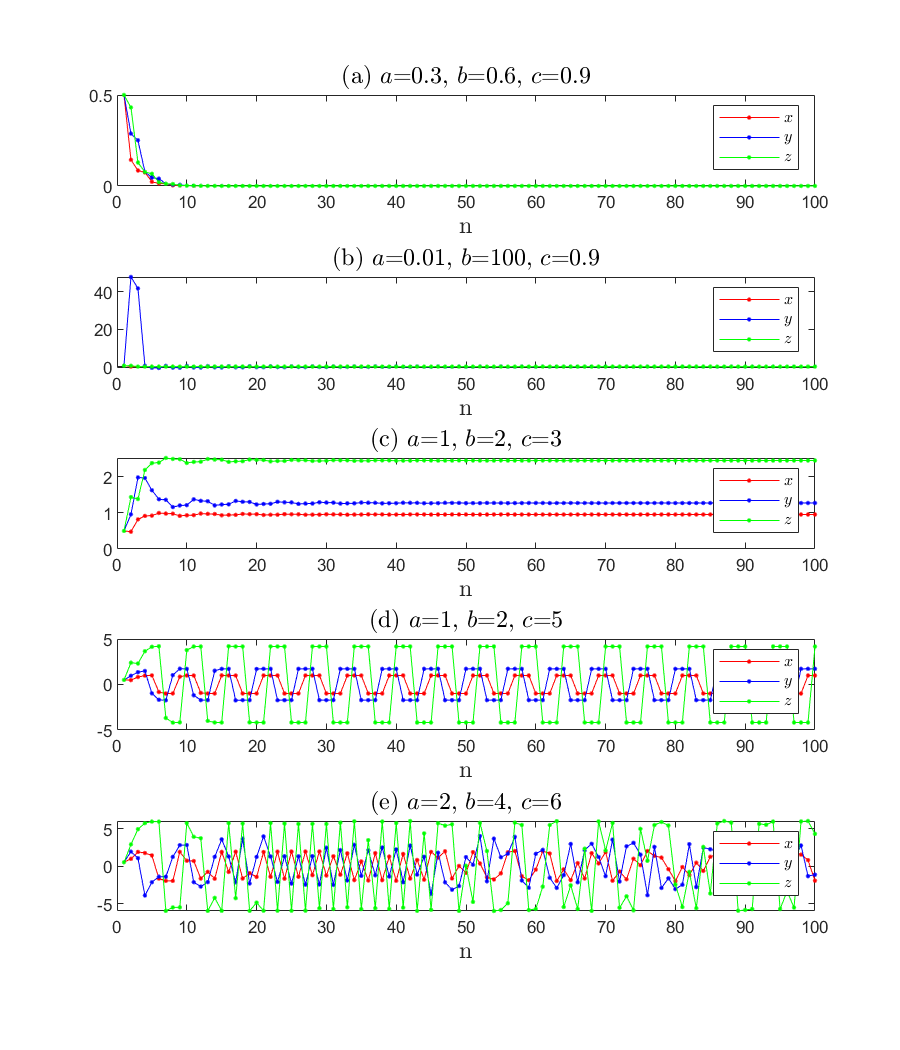}
\caption{3D-SCDS(\ref{eq:scs2}) transitions from stability to chaos}
\label{fig:SCSzhouqi}
\end{figure}

We have respectively plotted the time-domain diagrams of the system for different values of $a$, $b$ and $c$, as shown in Fig.\ref{fig:SCSzhouqi}. As $a$, $b$ and $c$ increase, the system sequentially experiences convergence to 0 (Fig.\ref{fig:SCSzhouqi}(a)(b)), convergence to a non-zero fixed point (Fig.\ref{fig:SCSzhouqi}(c)), a periodic state (Fig.\ref{fig:SCSzhouqi}(d)), and a chaotic state (Fig.\ref{fig:SCSzhouqi}(e)). Furthermore, as seen from the Fig.\ref{fig:SCSzhouqi} (b), regardless of how much the parameters differ, as long as the condition $abc<1$ is satisfied, the system remains stable.

\begin{figure}
\centering
\includegraphics[width=1\linewidth]{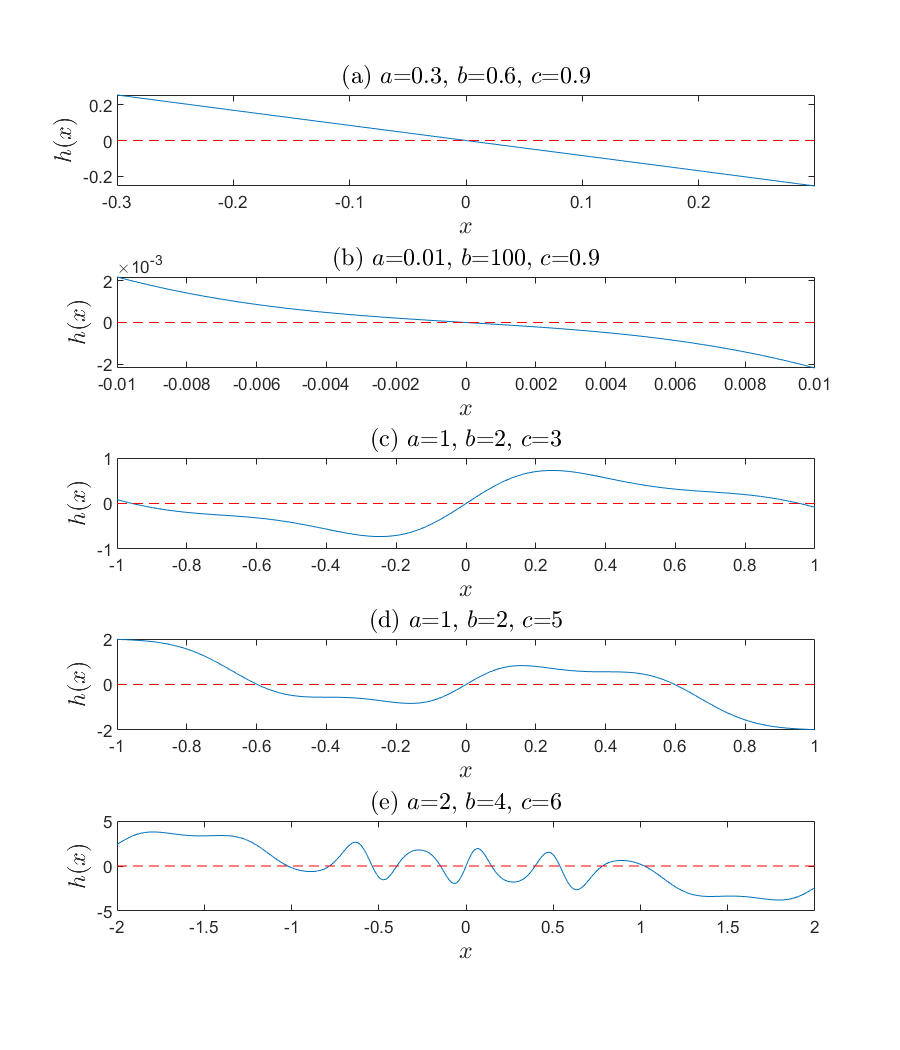}
\caption{Fixed points of 3D-SCDS(\ref{eq:scs2}) with different parameters}
\label{fig:SCSdzqfixed}
\end{figure}

By solving for the system's fixed points under the same parameters as in Fig.\ref{fig:SCSzhouqi}, we obtain Fig.\ref{fig:SCSdzqfixed}. 
It can be observed that when $\left| abc \right| < 1$, the system's time-domain diagram stabilizes at zero and there is only one fixed point. When the system is in the periodic or chaotic states, there are multiple fixed points. Therefore, the results of simulation experiments are completely consistent with Theorem 1.


The above theorem is extended to $m$-dimensional system, and the following corollary can be obtained.

\textbf{Remark 1:} For $m$-dimensional SCDS(\ref{eq:scs}), if $\left| {\alpha_1 \alpha_2 \ldots \alpha_m}\right|<1$, then the fixed points are stable.

\subsection{\label{sec:level2} Multidimensional Chebyshev System in real Field}

Inspired by the Chebyshev map and the coupling effect of sine and consine functions, we generalize it to $m$ dimensions and  it is essentially also a sine-cosine nonlinear system which  belongs to SCNSF(\ref{eq:sccs}). Its mathematical equation can be described as follows
\begin{eqnarray}
\left\{ \begin{array}{l}
	x_{1,(n + 1)} = g\big({\beta_1}f(x_{2,(n)})\big),\\
	x_{2,(n + 1)} = g\big({\beta_2}f(x_{3,(n)})\big),\\
	......\\
	x_{{m - 1},(n + 1)} = g\big({\beta_{m - 1}}f(x_{m,(n)})\big),\\
	x_{m,(n + 1)} = g\big({\beta_m}f(x_{1,(n)})\big),
\end{array} \right.\
\label{eq:cs}
\end{eqnarray}
where $g$ is a sine or cosine function and $f$ is arcsine and arccosine function. It is also a speical case of Eq.(\ref{eq:sccs}) with $\alpha_1=\alpha_2= \ldots=\alpha_m=1$, $\gamma_1=\gamma_2=\ldots=\gamma_m=0$.

We take $m=3$ as an exmple and obtain a three-dimensional Chebyshev system(3D-CS). Its expression is as follows:

\begin{eqnarray}
\left\{ \begin{array}{l}
	{x_{(n + 1)}} = \sin (a{\sin ^{ - 1}}({y_{(n)}})),\\
	{y_{(n + 1)}} = \sin (b{\sin ^{ - 1}}({z_{(n)}})),\\
	{z_{(n + 1)}} = \sin (c{\sin ^{ - 1}}({x_{(n)}})).
\end{array} \right.
\label{eq:3Chebyshev}
\end{eqnarray}

\begin{figure}
\centering
\includegraphics[width=0.9\linewidth]{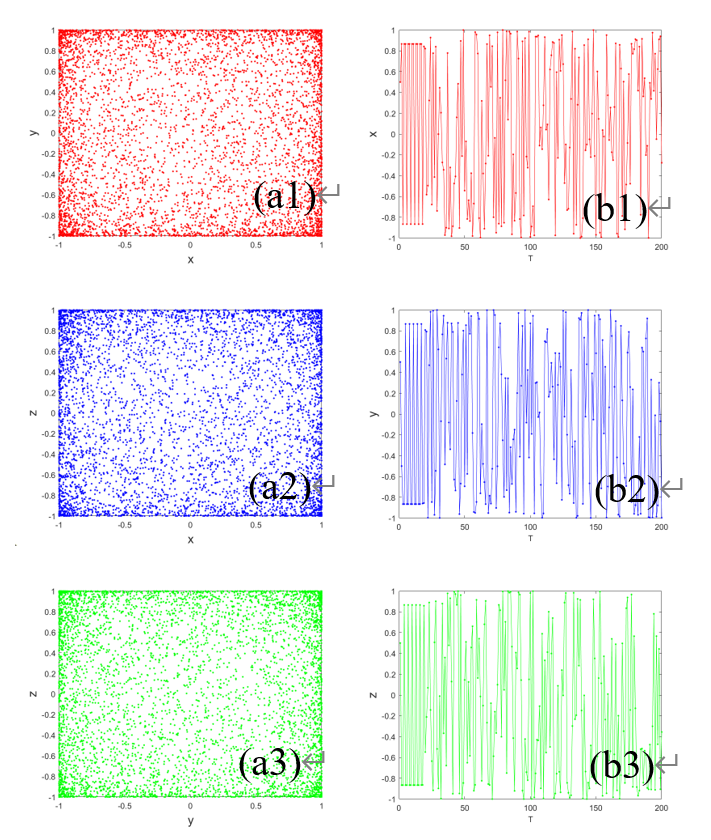}
\caption{The phase portrait and time-domain diagram of 3D-CS(\ref{eq:3Chebyshev}). (a1) $x-y$ (a2) $x-z$ (a3) $y-z$  (b1) $t-x$ (b2) $t-y$ (b3) $t-z$.}
\label{fig:sx2}
\end{figure}

\begin{figure}
\centering
\includegraphics[width=0.8\linewidth]{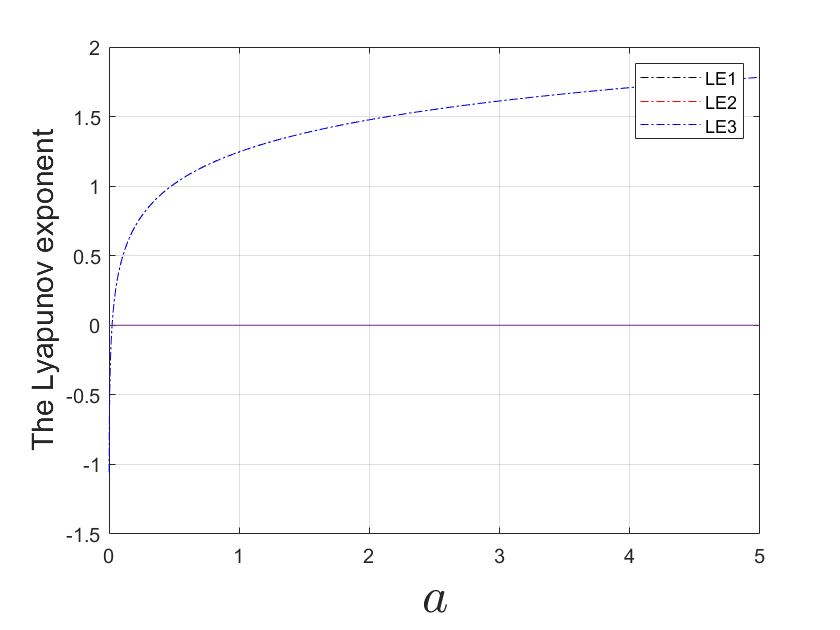}
\caption{Lyapunov exponents of 3D-CS(\ref{eq:3Chebyshev})}
\label{fig:Lyapunov exponents of CS}
\end{figure}

\begin{figure}
\centering
\includegraphics[width=0.8\linewidth]{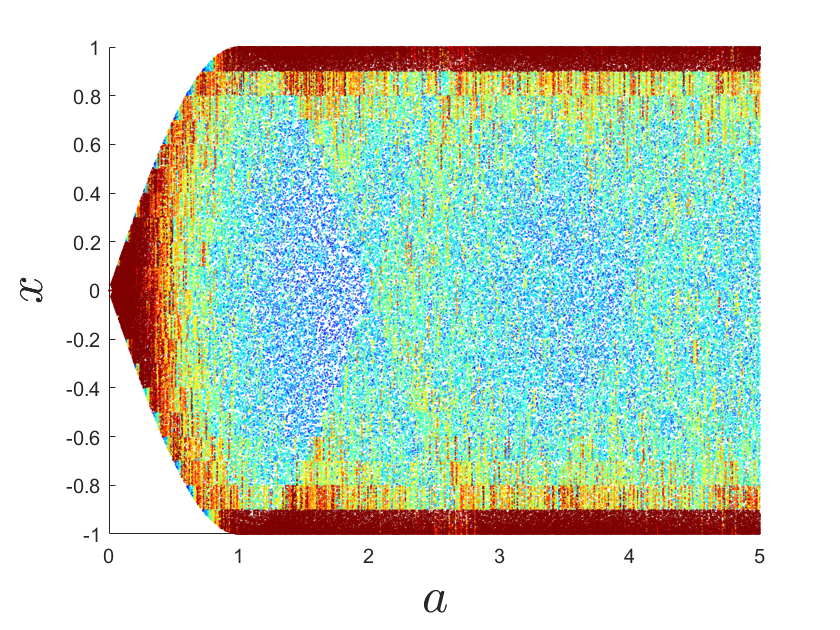}
\caption{The bifurcation diagram of 3D-CS(\ref{eq:3Chebyshev})}
\label{fig:CS BifurcationGragh}
\end{figure}

When $a = 4$, $b = 7$ and $c = 10$, the phase portrait and time-
domain diagram of the 3D-CS (\ref{eq:3Chebyshev}) can be depicted in Fig.\ref{fig:sx2}.
The LEs are given in Fig.\ref{fig:Lyapunov exponents of CS}, and the bifurcation diagram of system is depicted in Fig.\ref{fig:CS BifurcationGragh}.

From Fig.\ref{fig:Lyapunov exponents of CS} and Fig.\ref{fig:CS BifurcationGragh},  there are positive LE and the chaotic range of 3D-CS gradually expands as the parameter $a$ increases, which leads to the conclusion that the system is chaotic. 

Similarly, let us compute the fixed points of the 3D-CS, i.e.
\begin{eqnarray}
\left\{ \begin{array}{l}
	x_f = \sin (a{\sin ^{ - 1}}y_f),\\
	y_f = \sin (b{\sin ^{ - 1}}z_f),\\
	z_f = \sin (c{\sin ^{ - 1}}x_f),
\end{array} \right.
\label{eq:ccs}
\end{eqnarray}
Simplify to equations involving only $x_f$
\begin{eqnarray}
x_f = \sin(a \sin^{-1}(\sin(b \sin^{-1}(\sin(c \sin^{-1}x_f))))),
\end{eqnarray}
Obtain the formula for solving the fixed points
\begin{align}
	h(x_f) &= \sin\left(a \sin^{-1}\left(\sin\left(b \sin^{-1}\left(\sin\left(c \sin^{-1}x_f\right)\right)\right)\right)\right) \notag \\
	&\quad - x_f.
\end{align}

\begin{figure}
\centering
\includegraphics[width=0.8\linewidth]{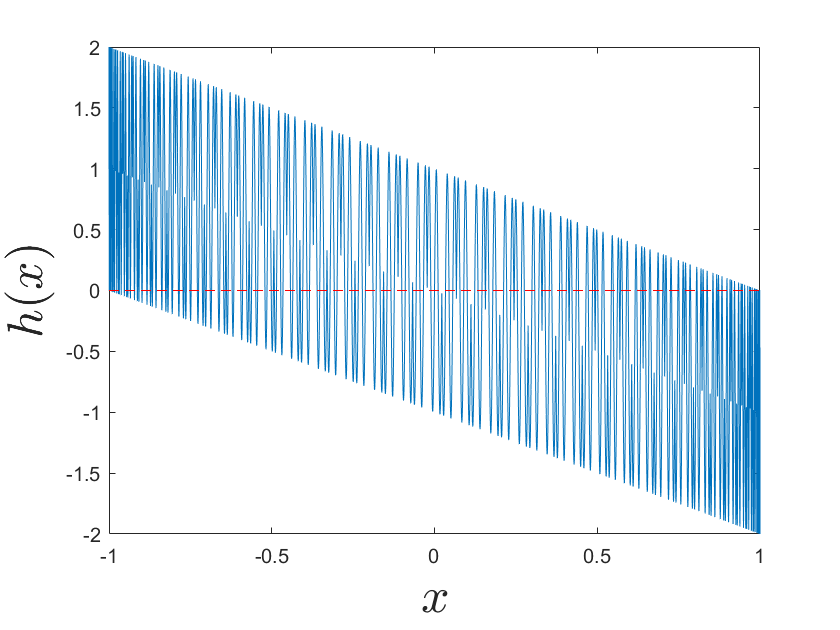}
\caption{Fixed points of 3D-CS(\ref{eq:3Chebyshev})}
\label{fig:CS FixedPoint}
\end{figure}

The fixed points of 3D-CS are plotted in Fig.\ref{fig:CS FixedPoint}. Obviously, there are multiple fixed points when $a=4$, $b=7$ and $c=10$, which is similar to those in 3D-SCDS.

We obtian the Jacobian matrix $J$, and get
\begin{eqnarray}
J = {\left. {\left[ {\begin{array}{*{20}{c}}
				0&{\frac{{a\cos (a{{\sin }^{ - 1}}y)}}{{\sqrt {1 - {y^2}} }}}&0\\
				0&0&{\frac{{b\cos (b{{\sin }^{ - 1}}z)}}{{\sqrt {1 - {z^2}} }}}\\
				{\frac{{c\cos (c{{\sin }^{ - 1}}x)}}{{\sqrt {1 - {x^2}} }}}&0&0
		\end{array}} \right]} \right|_{({x_f},{y_f},{z_f})}},
\label{fig:SCSJ}
\end{eqnarray}

We solve the eigenvalues at fixed points, and obtain
\begin{eqnarray}
{\lambda _{1,2,3}} = \sqrt[3]{{ \pm abc}}.
\label{fig:SCSJ1}
\end{eqnarray}

It is evident that condition Eq.(\ref{fig:SCSJ1}) is also a sufficient condition for the stability of the fixed points in the 3D-CS(\ref{eq:3Chebyshev}).

\textbf{Theorem 2:} For 3D-CS(\ref{eq:3Chebyshev}), if $\left| abc \right| < 1$, then the fixed points are stable.

\begin{figure}
\centering
\includegraphics[width=1\linewidth]{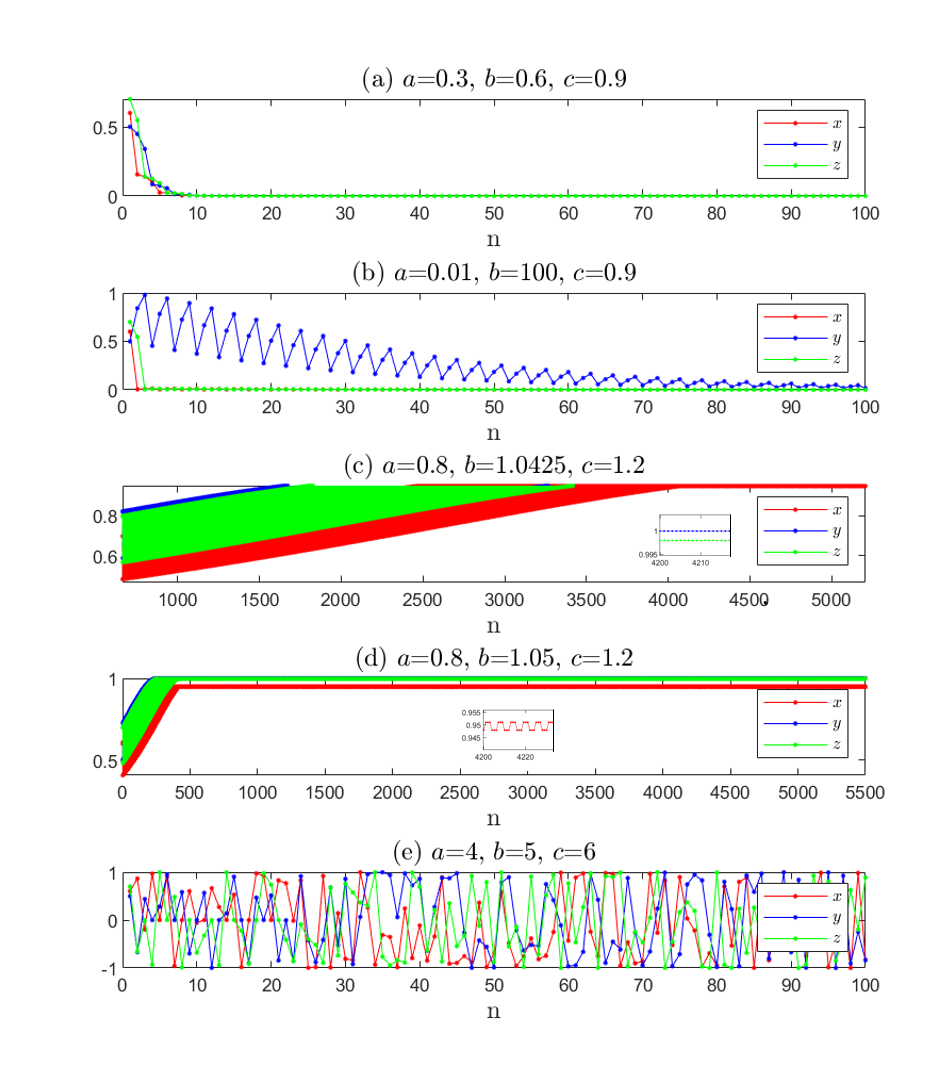}
\caption{3D-CS(\ref{eq:3Chebyshev}) transitions from stability to chaos}
\label{fig:CSzhouqi}
\end{figure}

We respectively plot the time-domain diagrams of 3D-CS with the increase of $a$, $b$ and $c$, as shown in Fig.\ref{fig:CSzhouqi}. The system undergoes a sequential transition from convergence to zero Fig.\ref{fig:CSzhouqi}(a)(b), converge to a non-zero fixed point Fig.\ref{fig:CSzhouqi}(c), to a periodic state Fig.\ref{fig:CSzhouqi}(d), and eventually to a chaotic state Fig.\ref{fig:CSzhouqi}(e). Fig.\ref{fig:CSzhouqi}(b) demonstrates that the condition remains applicable even when the parameters differ greatly, such as a=0.01,b=100.
When $a$=0.8, $b$=1.0425, $c$=1.2, the system requires more than 3800 iterations to stabilize at a non-zero fixed point depicted in Fig.\ref{fig:CSzhouqi}(c). However, when $b$ experiences a slight increase ($b$=1.05), the system exhibits a periodic state. Consequently, it becomes rather challenging to identify the parameters that can make the system converge to a non-zero fixed point. This implies that even a minor alteration in the values of parameters can lead to a significant change in the system's behavior and stability characteristics, highlighting the sensitivity of the system to parameter variations.

\begin{figure}
\centering
\includegraphics[width=1\linewidth]{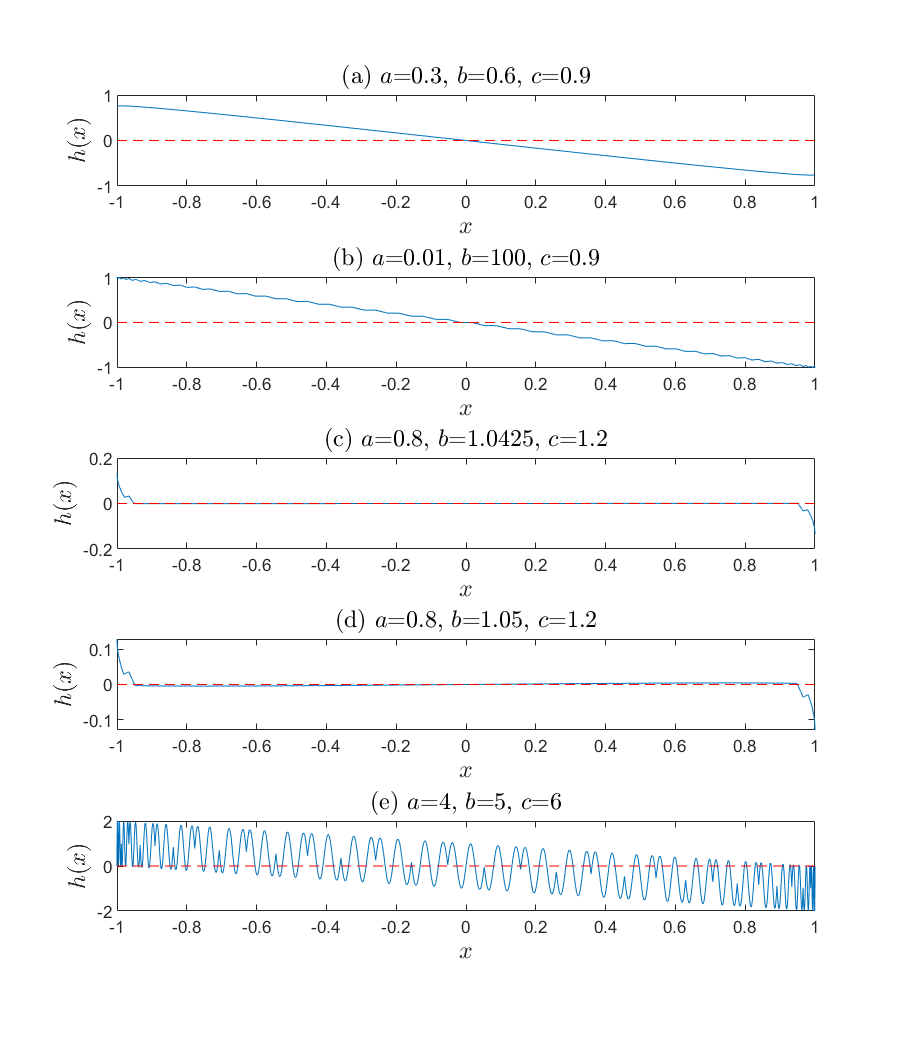}
\caption{Fixed points of 3D-CS(\ref{eq:3Chebyshev}) with different parameters}
\label{fig:CSdzqfixed}
\end{figure}

Solving for the system's fixed points under the same parameters as in Fig.\ref{fig:CSzhouqi}, we obtain Fig.\ref{fig:CSdzqfixed}. It can be observed that when $\left| abc \right| < 1$, the system's time-domain diagram stabilizes at zero. When $\left| abc \right| \geq 1$, the system is in the periodic or chaotic states, there are multiple fixed points. Therefore, the results of simulation experiments are completely consistent with Theorem 2.

Similiarly, we extend Theorem 2 to $m$-dimensional CS(\ref{eq:cs}), and the following corollary can be obtained.

\textbf{Remark 2:} For $m$-dimensional CS(\ref{eq:cs}), if $\left| {\alpha_1 \alpha_2 \ldots \alpha_m}\right|<1$, then the fixed points are stable.

\subsection{\label{sec:level2} Sine-Logistic System in real Field}

Inspired by the Logistic function and the sine function, we propose the following $m$-dimensional Sine-Logistic System.
\begin{eqnarray}
\left\{ \begin{array}{l}
	x_{1,(n + 1)} = {\alpha_1}g\big(f(x_{2,(n)})\big),\\
	x_{2,(n + 1)} = {\alpha_2}g\big(f(x_{3,(n)})\big),\\
	......\\
	x_{{m - 1},(n + 1)} = {\alpha_{m - 1}}g\big(f(x_{m,(n)})\big),\\
	x_{m,(n + 1)} = {\alpha_m}g\big(f(x_{1,(n)})\big),
\end{array} \right.\
\label{sls}
\end{eqnarray}
where $g$ is the Logistic mapping and $f$ is Sine or Cosine function. It is also a speical case of SCNSF(\ref{eq:sccs}) with $\beta_1=\beta_2= \ldots=\beta_m=1$, $\gamma_1=\gamma_2=\ldots=\gamma_m=0$.

With $m=3$, the three-dimensional Sine-Logistic System(3D-SLS) is expressed as follows:
\begin{eqnarray}
\left\{ \begin{array}{l}
	{x_{(n + 1)}} = a\sin ({y_{(n)}})(1 - \sin ({y_{(n)}})),\\
	{y_{(n + 1)}} = b\sin ({z_{(n)}})(1 - \sin ({z_{(n)}})),\\
	{z_{(n + 1)}} = c\sin ({x_{(n)}})(1 - \sin ({x_{(n)}})),
\end{array} \right.
\label{eq: slccs}
\end{eqnarray}
\begin{figure}
\centering
\includegraphics[width=0.9\linewidth]{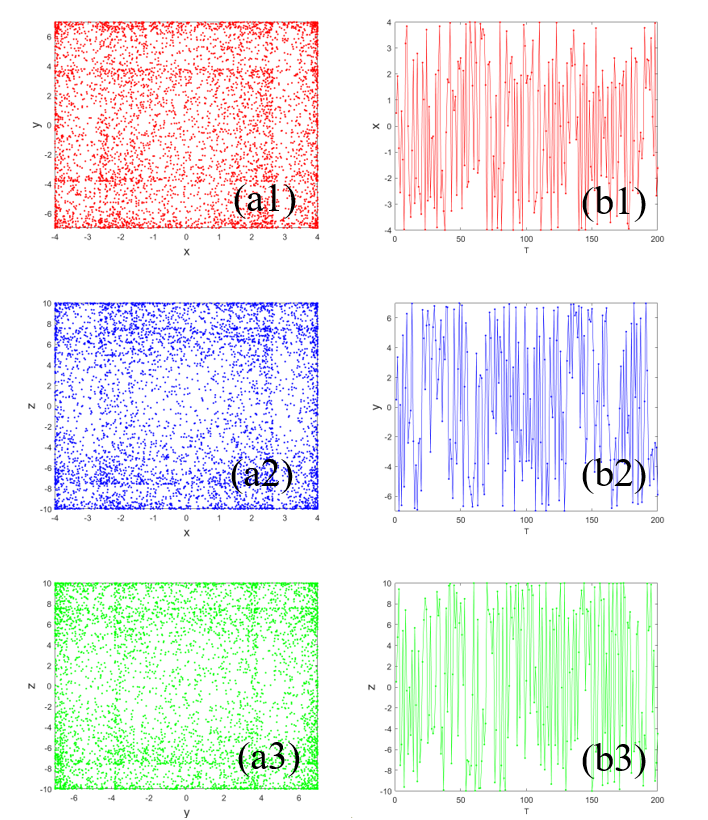}
\caption{The phase portrait and time-domain diagram of 3D-SLS (\ref{eq: slccs}). (a1) $x-y$ (a2) $x-z$ (a3) $y-z$  (b1) $t-x$ (b2) $t-y$ (b3) $t-z$ .}
\label{fig: slccs}
\end{figure}

\begin{figure}
\centering
\includegraphics[width=0.8\linewidth]{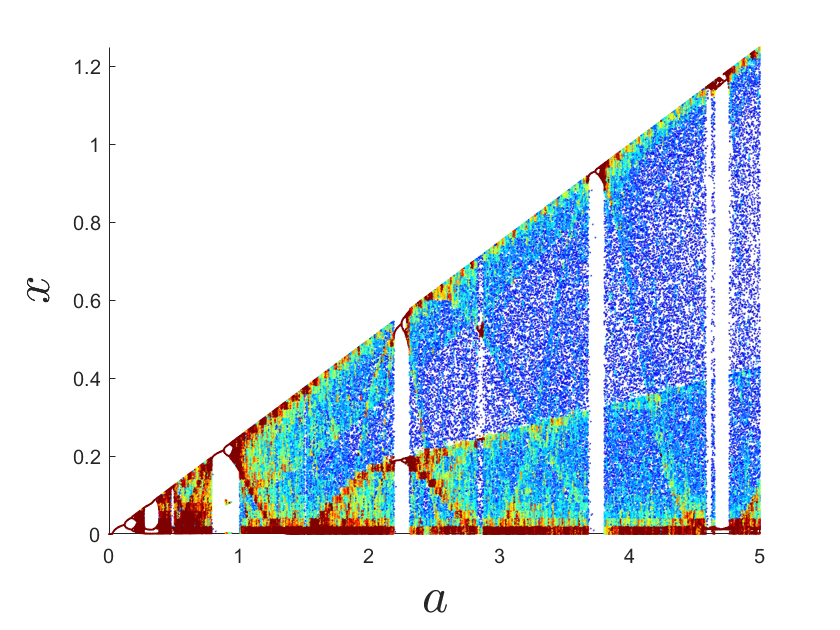}
\caption{The bifurcation diagram of 3D-SLS(\ref{eq: slccs})}
\label{fig:SLCCS BifurcationGragh}
\end{figure}

\begin{figure}
\centering
\includegraphics[width=0.8\linewidth]{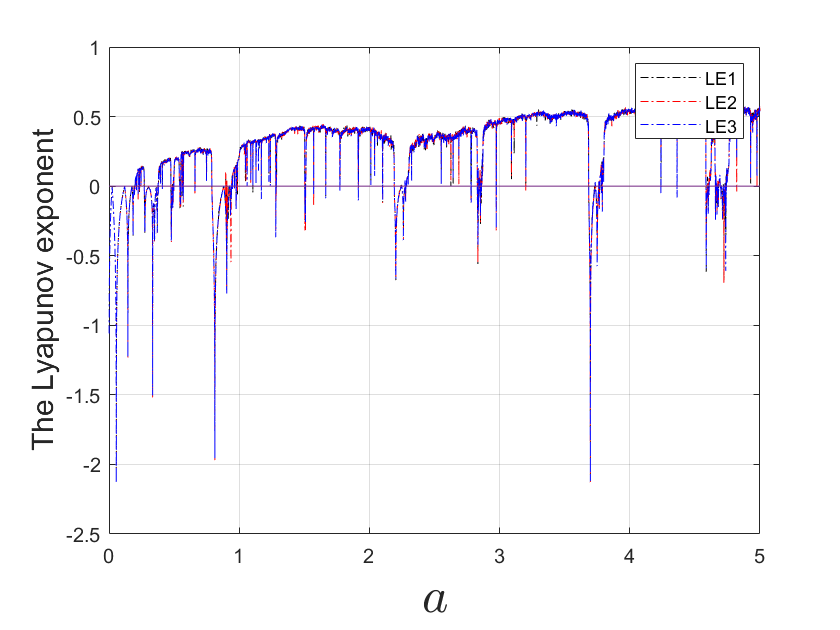}
\caption{Lyapunov exponents of 3D-SLS(\ref{eq: slccs})}
\label{fig:Lyapunov exponents of SLCCS}
\end{figure}

When $a=5$, $b=6$ and $c=7$, the phase portrait and time-domain plot generated by 3D-SLS are depicted as Fig.\ref{fig: slccs}. The bifurcation diagram of the 3D-SLS is plotted as $0\le a<5$, with $b=6$ and $c=7$, as shown in 
Fig.\ref{fig:SLCCS BifurcationGragh}. The LEs are depicted in Fig.\ref{fig:Lyapunov exponents of SLCCS}. From the above figures, there are positive LE and the chaotic range of 3D-SLS gradually expands as the parameter $a$ increases, which indicates the 3D-SLS (\ref{eq: slccs}) is chaotic.

As for the fixed points of Eq.(\ref{eq: slccs}),  we obtain
\begin{eqnarray}
\left\{ \begin{array}{l}
	x_f = a\sin y_f(1 - \sin y_f),\\
	y_f = b\sin z_f(1 - \sin z_f),\\
	z_f = b\sin x_f(1 - \sin x_f),
\end{array} \right.
\label{eq: slccsfixed}
\end{eqnarray}

We can simplify  Eq.(\ref{eq: slccsfixed}) as follows:

\begin{equation}
\begin{split}
	x_f = & \, a \sin \left( b \sin \left( c \sin x_f ( 1 - \sin x_f ) \right) \right) \\
	& \times \left( 1 - \sin \left( c \sin x_f ( 1 - \sin x_f ) \right) \right) \\
	& \times \left( 1 - \sin \left( b \sin \left( c \sin x_f ( 1 - \sin x_f ) \right) \right) \right) \\
	& \times \left( 1 - \sin \left( c \sin x_f ( 1 - \sin x_f ) \right) \right),
\end{split}
\label{eq: slccsfixed1}
\end{equation}

In order to solve the fixed points, we can rewrite Eq.(\ref{eq: slccsfixed1}) in the following form Eq.(\ref{eq: slccsfixed2}).
\begin{equation}
\begin{split}
	h(x_f) = & \, x_f - a \sin \left( b \sin \left( c \sin x_f ( 1 - \sin x_f ) \right) \right) \\
	& \times \left( 1 - \sin \left( c \sin x_f ( 1 - \sin x_f ) \right) \right) \\
	& \times \left( 1 - \sin \left( b \sin \left( c \sin x_f ( 1 - \sin x_f ) \right) \right) \right) \\
	& \times \left( 1 - \sin \left( c \sin x_f ( 1 - \sin x_f ) \right) \right),
\end{split}
\label{eq: slccsfixed2}
\end{equation}

\begin{figure}
\centering
\includegraphics[width=0.8\linewidth]{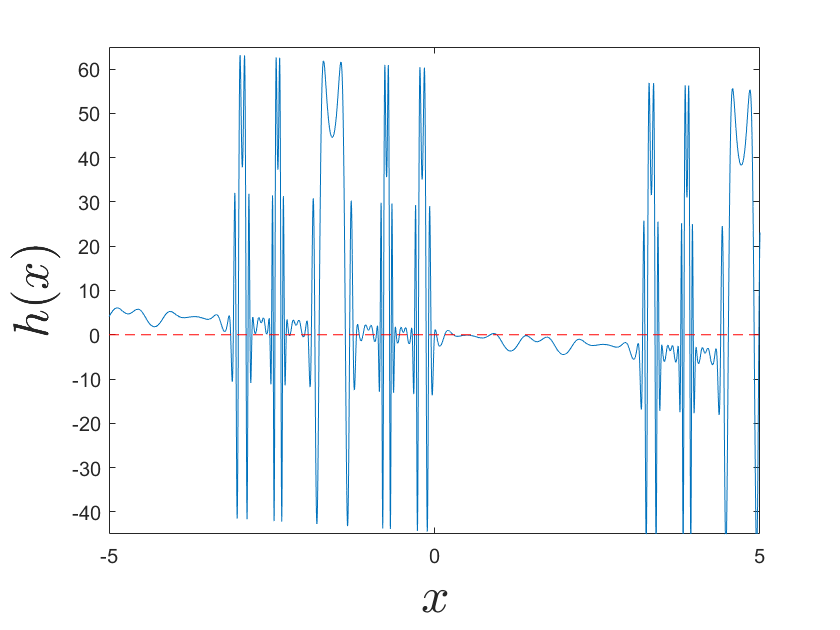}
\caption{Fixed points of SLS(\ref{eq: slccs})}
\label{fig:SLCCS FixedPoint}
\end{figure}

We obtain the following Jacobian matrix $J$.
\begin{equation}
J = \left. \left[ \begin{smallmatrix}
	0 & a \cos y (1 - 2 \sin y) & 0 \\
	0 & 0 & b \cos z (1 - 2 \sin z) \\
	c \cos x (1 - 2 \sin x) & 0 & 0
\end{smallmatrix} \right] \right|_{(x_f, y_f, z_f)}
\end{equation}

and calculate its eigenvalue
\begin{eqnarray}
\lambda^3 &=& abc\cos x_f\cos y_f\cos z_f \nonumber \\
&& \times(2\sin x_f - 1)(2\sin y_f - 1)(2\sin z_f - 1),
\label{fig:SLSJ1}
\end{eqnarray}
Due to the fact that $\left|{\cos x_f} \right| \le 1$,$\left|2\sin x_f - 1\right| \le 3$,
\begin{eqnarray}
{\lambda ^3} \in [ - 27abc,27abc],
\label{fig:SLSJ2}
\end{eqnarray}
Then
\begin{eqnarray}
\left| {abc} \right| < \frac{1}{27}.
\label{fig:SLSJ3}
\end{eqnarray}

Eq.(\ref{fig:SLSJ3}) provides a sufficient condition for the stability of the system's fixed point, leading to the following theorem.

\textbf{Theorem 3:} For the 3D-SLS(\ref{eq: slccs}), if $\left| abc \right| < 1/27$, then the fixed points are stable.

\begin{figure}
\centering
\includegraphics[width=1\linewidth]{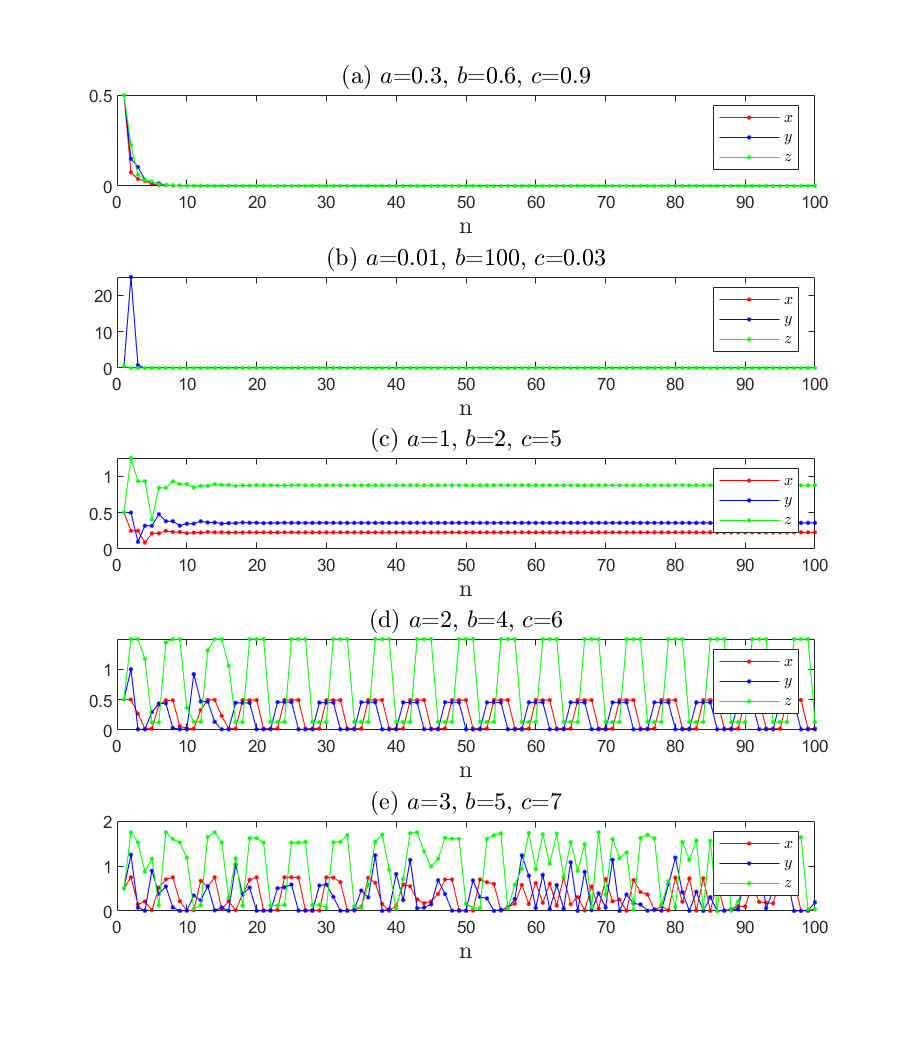}
\caption{3D-SLS(\ref{eq: slccs}) transitions from stability to chaos}
\label{fig:SLSzhouqi}
\end{figure}

We respectively plot the time-domain diagrams of 3D-SLS with the increase of $a$, $b$ and $c$, as shown in Fig.\ref{fig:SLSzhouqi}. The system undergoes a sequence of states first converging to zero Fig.\ref{fig:SLSzhouqi}(a)(b), then to a non-zero fixed point Fig.\ref{fig:SLSzhouqi}(c), followed by a periodic state Fig.\ref{fig:SLSzhouqi}(d), and finally reaching a chaotic state Fig.\ref{fig:SLSzhouqi}(e).

\begin{figure}
\centering
\includegraphics[width=1\linewidth]{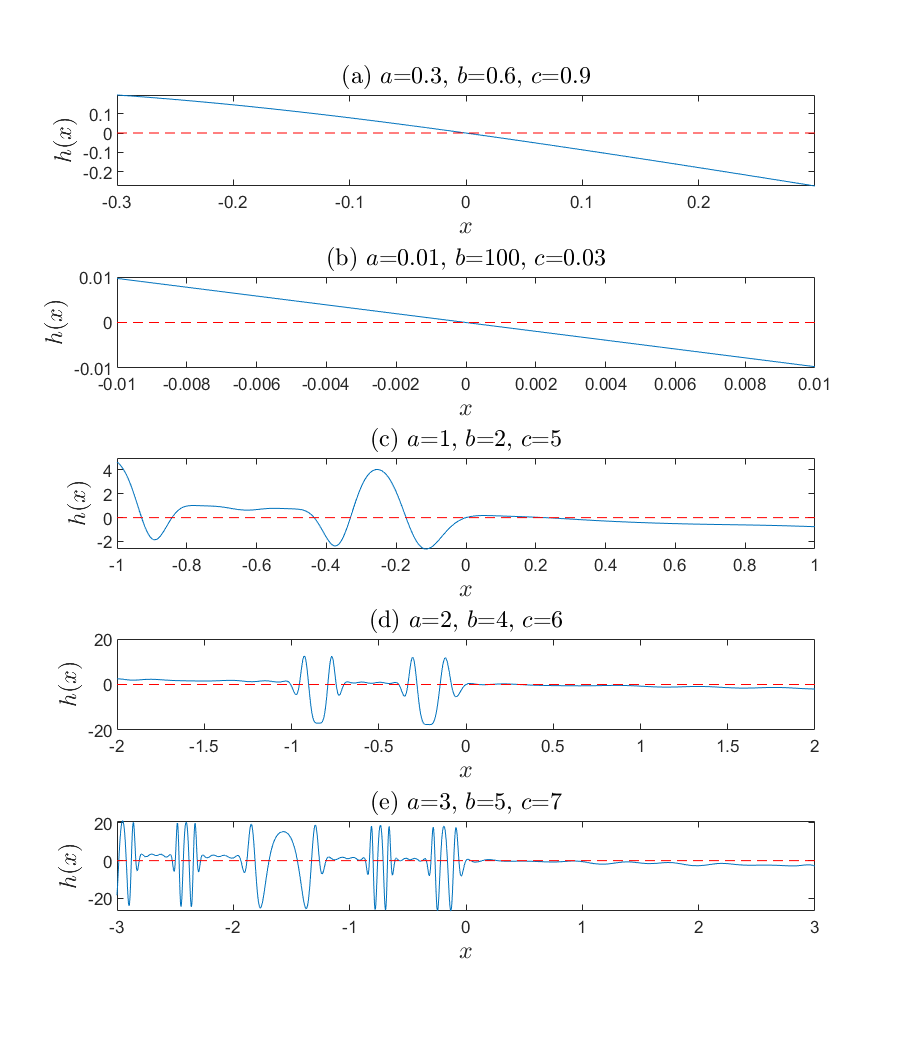}
\caption{Fixed points of 3D-SLS(\ref{eq: slccs}) with different parameters}
\label{fig:SLSdzqfixed}
\end{figure}

Similiarly, we obtain the fixed points of 3D-SLS with the same parameters as in Fig.\ref{fig:SLSzhouqi}, shown as Fig.\ref{fig:SLSdzqfixed}. When $\left| abc \right| < 1/27$, the system's time-domain diagram stabilizes at zero, which is consistent with Theorem 3.

\subsection{\label{sec:level2}MCU-based hardware implementation}

The digital hardware implementation platform is a hardware platform with digital circuits and a microcontroller (MCU). The MCU offers the advantages of high speed and strong flexibility. Compared to traditional analog circuit, the digital hardware
platform can realize complex algorithms, such as nonlinear sine and cosine functions. 
\begin{figure}
\centering
\includegraphics[width=0.8\linewidth]{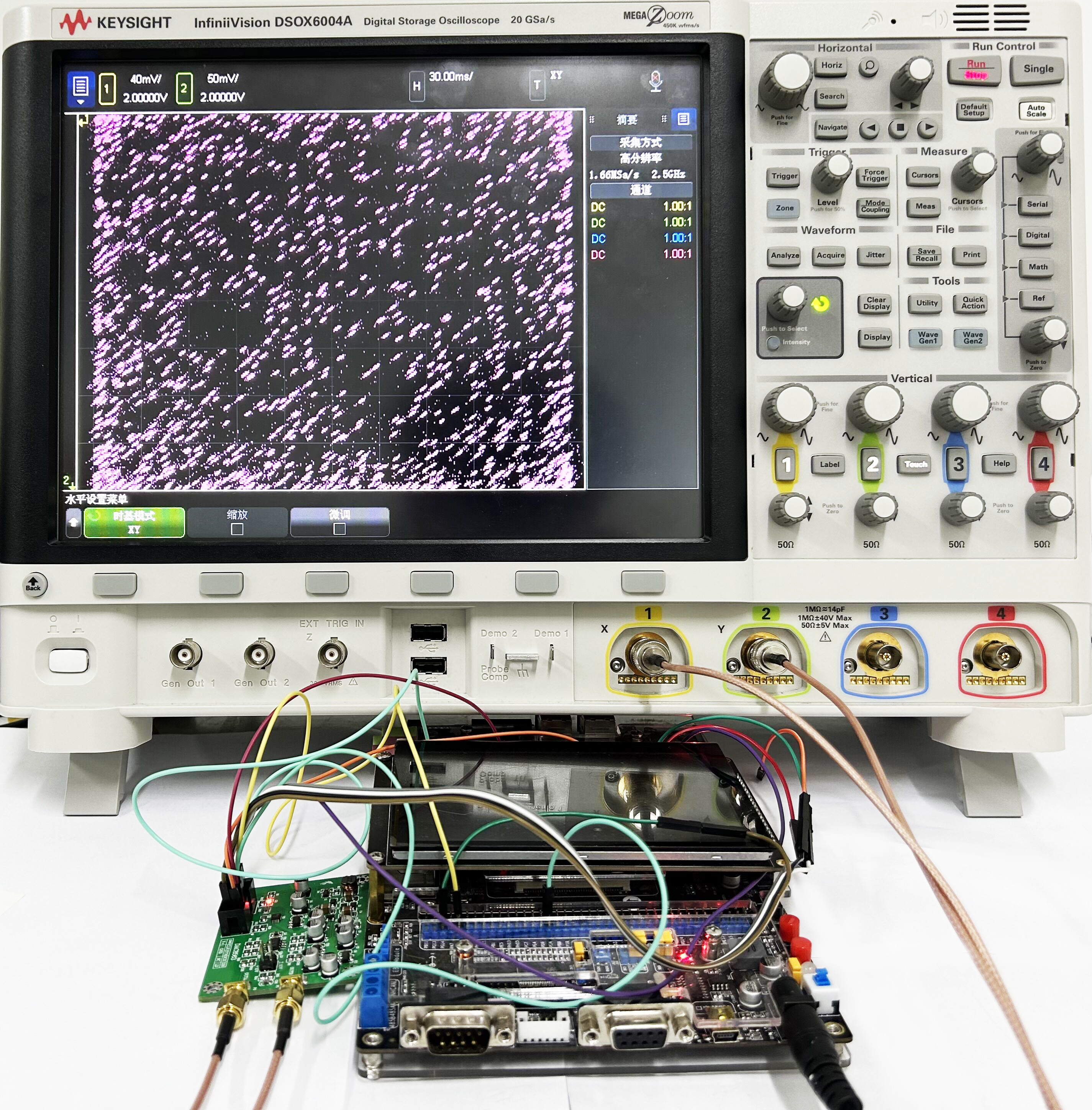}
\caption{The microcontroller development board is connected to a digital oscilloscope for experimental testing and results.}
\label{fig:environment}
\end{figure}

The proposed MCU-based digital hardware platform consists of a 32-bit STM32H743IIT6 microcontroller, a 16-bit DAC8563 D/A converter, and other peripheral voltage conversion circuits. The MCU is used to construct the SCNSF while the D/A converter generates multi-channel voltage sequences. The initial values and parameters of the SCNSF are configured using C language, and the execution code is compiled into the MCU. The digital hardware platform generates the voltage sequences of the SCNSF with the results displayed on a digital oscilloscope, as shown in Fig.\ref{fig:environment}.

\begin{figure}
\centering
\includegraphics[width=0.7\linewidth]{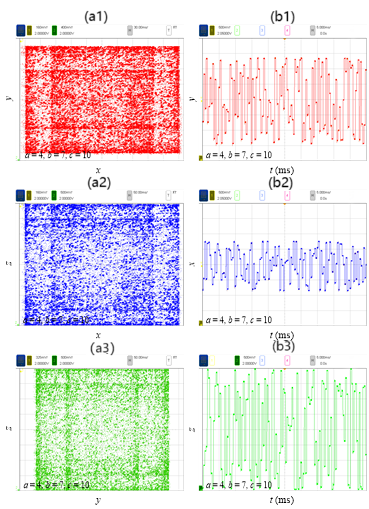}
\caption{The phase portrait and time-domain diagram of 3D-SCS with $a=4, b=7, c=10$  (a1) $x-y$ (a2) $x-z$ (a3) $y-z$  (b1) $t-x$ (b2) $t-y$ (b3) $t-z$ .}
\label{fig:sccs}
\end{figure}

\begin{figure}
\centering
\includegraphics[width=0.8\linewidth]{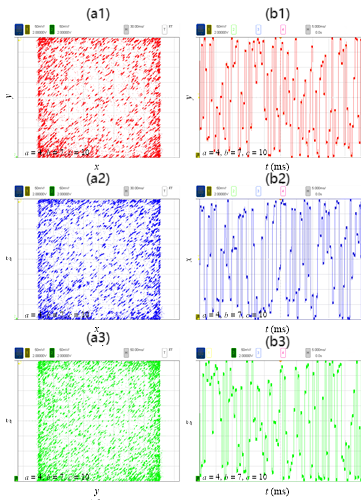}
\caption{ The phase portrait and time-domain diagram of 3D-CS with $a=4, b=7, c=10$ (a1) $x-y$ (a2) $x-z$ (a3) $y-z$  (b1) $t-x$ (b2) $t-y$ (b3) $t-z$ .}
\label{fig:3dccs}
\end{figure}

\begin{figure}
\centering
\includegraphics[width=0.8\linewidth]{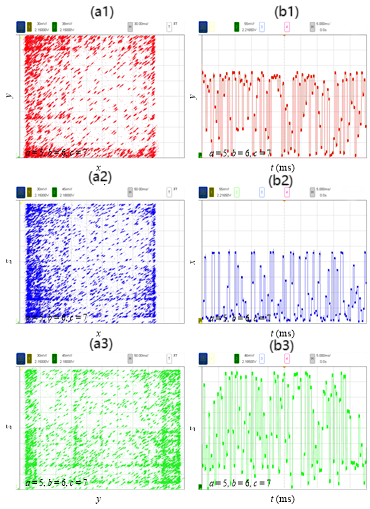}
\caption{ The phase portrait and time-domain diagram of 3D-SLS with $a=5, b=6, c=7$ (a1) $x-y$ (a2) $x-z$ (a3) $y-z$  (b1) $t-x$ (b2) $t-y$ (b3) $t-z$ .}
\label{fig:slccs}
\end{figure}

According to the equations of 3D-SCDS, 3D-CS and 3D-SLS, the corresponding system parameters and initial conditions are set. The experimental test results of the MCU-based digital hardware platform are shown in Fig.\ref{fig:sccs}, Fig.\ref{fig:3dccs} and Fig.\ref{fig:slccs}. The physical experimental results closely match the numerical simulations Fig.\ref{fig:sx1}, Fig.\ref{fig:sx2} and Fig.\ref{fig: slccs}, demonstrating that the proposed hardware implementation can successfully realize the SCNSF.Therefore, the proposed SCNSF has a simple structure and can be easily implemented in hardware.

\subsection{\label{sec:level2}The mechanism of chaos generation from Sine-Cosine Nonlinear Family.}

For the above three types of Sin-Consine nonlinear systems which only have simple mappings, how do they generate complex chaotic phenomena? Here, starting from the sensitivity of chaos to initial values, we employ decoupling analysis and discover the mechanism by which SCNSF enters chaos.

\begin{figure}
\centering
\includegraphics[width=0.8\linewidth]{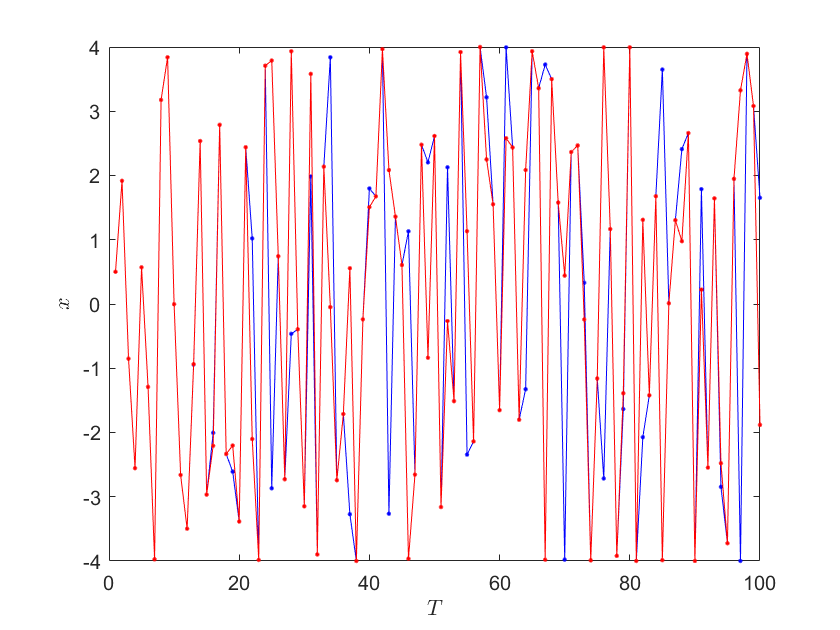}
\caption{Sensitivity analysis of the 3D-SCDS}
\label{fig:sccs1}
\end{figure}

When $a=4$, $b=7$ and $c=10$, the sensitivity diagrams of the 3D-SCDS are displayed in Fig.\ref{fig:sccs1} where the blue line corresponds to the initial value (0.5,0.5,0.5) and the red line corresponds to (0.500000001,0.5,0.5). It is evident that the 3D-SCDS exhibits significant sensitivity to initial conditions. 

\begin{table}
\caption{\label{tab:table2}Some values of $x$, $y$, and $z$ generated from 3D-SCDS}
\begin{ruledtabular}
	\fontsize{8pt}{10pt}\selectfont 
	\begin{tabular}{lcccccc}
		$n$ & 85 & 86 & 87 & 88 & 89 & 90 \\
		\hline
		\multicolumn{7}{c}{Initial values (0.5, 0.5, 0.5)} \\
		\hline
		$x$ & \textbf{3.648403} & 0.010416 & 1.304823 & \textbf{2.410634} & 2.660841 & -3.9993 \\ 
		$y$ & 3.138989 & -3.47388 & \textbf{6.930014} & 0.727776 & -1.55215 & \textbf{2.67846} \\ 
		$z$ & 9.944073 & \textbf{-4.85391} & 0.104156 & 9.648372 & \textbf{6.675834} & 4.624453 \\
		\hline
		\multicolumn{7}{c}{Initial values (0.500000001, 0.5, 0.5)} \\
		\hline
		$x$ & \textbf{-3.98746} & 0.010416 & 1.304823 & \textbf{0.977688} & 2.660841 & -3.9993 \\ 
		$y$ & 3.138989 & -3.47388 & \textbf{6.530109} & 0.727776 & -1.55215 & \textbf{6.338946} \\ 
		$z$ & 9.944073 & \textbf{7.485493} & 0.104156 & 9.648372 & \textbf{8.292071} & 4.624453 \\ 
	\end{tabular}
\end{ruledtabular}
\label{table: a}
\end{table}

We randomly select six sequences from six trajectories generated by two initial values as shown in Table \ref{table: a}. There is an interesting dynamic change rule for six sequences. When the initial value is (0.500000001, 0.5, 0.5), the value of $x$ when $n=85$ and $n=88$ is different from that the sequence generated by the initial values (0.5, 0.5, 0.5); the value of $z$ when $n=86$ and $n=89$ is different from that the sequence generated by the initial values (0.5, 0.5, 0.5); the the value of $y$ when $n=87$ and $n=90$ is different from that the sequence generated by the initial values (0.5, 0.5, 0.5). 

In Eq.(\ref{eq:scs2}), by extracting every second value from $x$, $y$ and $z$, and defining them as\\ 
$o_1={x_{(1)}, x_{(4)}, x_{(7)}, \dots, x_{(85)}, x_{(88)},\dots}$,\\ 
$o_2={x_{(2)}, x_{(5)}, x_{(8)}, \dots, x_{(86)}, x_{(89)},\dots}$,\\ 
$o_3={x_{(3)}, x_{(6)}, x_{(9)}, \dots, x_{(87)}, x_{(90)},\dots}$,\\ 
$p_1={y_{(1)}, y_{(4)}, y_{(7)}, \dots, y_{(85)}, y_{(88)},\dots}$,\\ 
$p_2={y_{(2)}, y_{(5)}, y_{(8)}, \dots, y_{(86)}, y_{(89)},\dots}$,\\
$p_3={y_{(3)}, y_{(6)}, y_{(9)}, \dots, y_{(87)}, y_{(90)},\dots}$,\\ 
$q_1={z_{(1)}, z_{(4)}, z_{(7)}, \dots, z_{(85)}, z_{(88)},\dots}$,\\ 
$q_2={z_{(2)}, z_{(5)}, z_{(8)}, \dots, z_{(86)}, z_{(89)},\dots}$,\\
$q_3={z_{(3)}, z_{(6)}, z_{(9)}, \dots, z_{(87)}, z_{(90)},\dots}$,\\  
Therefore, for $o$,$p$,$q$ can be equivalently transformed into
\begin{eqnarray}
\left\{ \begin{array}{l}
	o_{1,(n + 1)} = a\sin \{ b\sin [c\sin (o_{1,(n)})]\}, \\
	o_{2,(n + 1)} = a\sin \{ b\sin [c\sin (o_{2,(n)})]\}, \\
	o_{3,(n + 1)} = a\sin \{ b\sin [c\sin (o_{3,(n)})]\}, \\
	p_{1,(n + 1)} = b\sin \{ c\sin [a\sin (p_{1,(n)})]\}, \\
	p_{2,(n + 1)} = b\sin \{ c\sin [a\sin (p_{2,(n)})]\}, \\
	p_{3,(n + 1)} = b\sin \{ c\sin [a\sin (p_{3,(n)})]\}, \\
	q_{1,(n + 1)} = c\sin \{ a\sin [b\sin (q_{1,(n)})]\},\\
	q_{2,(n + 1)} = c\sin \{ a\sin [b\sin (q_{2,(n)})]\},\\
	q_{3,(n + 1)} = c\sin \{ a\sin [b\sin (q_{3,(n)})]\},\\
\end{array} \right.
\label{eq9}
\end{eqnarray}
and the initial conditions is 
\begin{eqnarray}
\left\{ \begin{array}{l}
	o_{2,(1)} = asin(p_{1,(1)}),\\
	p_{2,(1)} = bsin(q_{1,(1)}),\\
	q_{2,(1)} = csin(o_{1,(1)}),\\
	o_{3,(1)} = asin(p_{2,(1)}),\\ 
	p_{3,(1)} = bsin(q_{2,(1)}),\\
	q_{3,(1)} = csin(o_{2,(1)}),\\
\end{array} \right.
\label{eq.opq}
\end{eqnarray}

According to Eq.(\ref{eq9}), the original system (3) is equivalent to the discrete point superposition of nine systems in Eq.(\ref{eq9}). Although the nine variables seem to be decoupled in Eq.(\ref{eq9}), the initial values of sequences $o_2, p_2, q_2$, are coupled with $o_1, p_1, q_1$, and the initial values of $o_3, p_3, q_3$ are coupled with $o_2, p_2, q_2$.
As shown in Eq.(\ref{eq.opq}), $o_{1,(1)}$ determines $q_{2,(1)}$, and $q_{2,(1)}$ determines $p_{3,(1)}$.  Simiarly, $p_{1,(1)}$ determines $o_{2,(1)}$, and $o_{2,(1)}$ determines $q_{3,(1)}$. $p_{1,(1)}$ determines $p_{2,(1)}$, and $p_{2,(1)}$ determines $o_{3,(1)}$. Therefore, the iteration sequence can be written as 
\begin{multline}
	o_{1,(1)} \rightarrow q_{2,(1)} \rightarrow p_{3,(1)} \rightarrow o_{1,(2)} \rightarrow q_{2,(2)} \\
	\rightarrow p_{3,(2)} \rightarrow o_{1,(3)} \rightarrow q_{2,(3)} \rightarrow p_{3,(3)} \rightarrow \dots
\end{multline}

The iteration order is $ o_1 \rightarrow q_2 \rightarrow p_3 $ and cycle in turn. Therefore, this cyclic iteration resulting from the coupling effect between variables destroys the periodicity of the sine and cosine functions, thereby giving rise to chaotic phenomena.

In summary, chaos can be generated solely by the coupling of initial values in a 3D-SCDS, and similar behavior occurs in other systems that meet the definition of the Sine-Cosine Nonlinear System Family. This highlights the profound sensitivity of nonlinear systems to initial conditions.

\section{\label{sec:level1} Sine-Cosine Nonlinear System Family in Complex Field}
When $x_1,x_2,\ldots,x_m$ are complex state variables, the SCNSF exhibits rich and colorful fractal phenomena.

\subsection{\label{sec:level2} SCDS in Complex Field}

For easy understanding, we take $m=1$ for SCDS as examples and discuss their dynamics characteristics in detail. 

The system equation of one-dimensional SCDS  is:

\begin{eqnarray}
{x_{(n + 1)}} = \alpha_1\sin ({x_{(n)}}),
\end{eqnarray}

\begin{figure}
\centering
\includegraphics[width=0.9\linewidth]{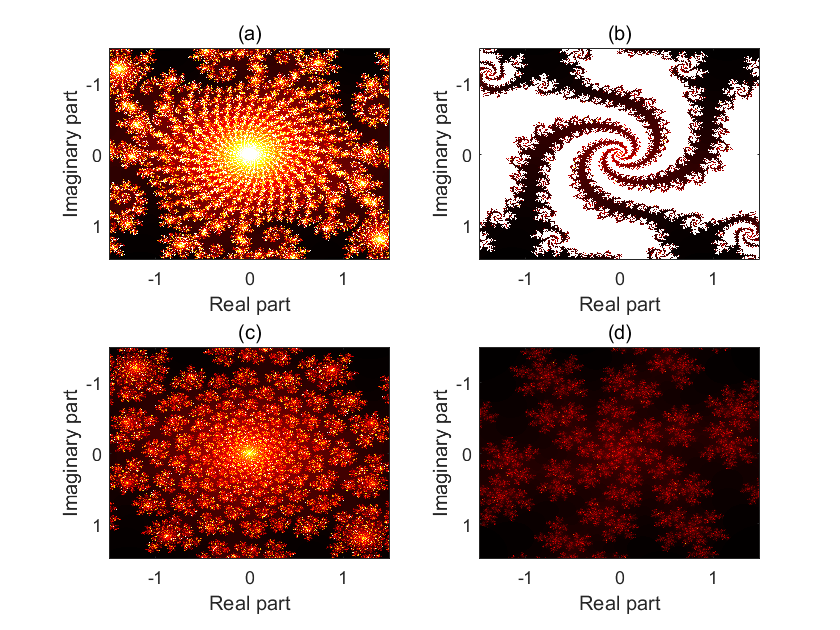}
\caption{Julia set of one-dimensional Sine-Cosine System
	(a)$\alpha_1=0.1+1.01i$
	(b)$\alpha_1=0.1+1.05i$
	(c)$\alpha_1=0.2+1.01i$
	(d)$\alpha_1=0.5+1.05i$}
\label{fig:1dSCS}
\end{figure}

We obtain the Julia set with different $\alpha_1$ depicted in Fig.\ref{fig:1dSCS}. When $\alpha_1=0.1+1.01i$, we observe a radiating structure composed of multiple recursive mini-fractals, exhibiting pronounced self-similarity in Fig.\ref{fig:1dSCS}(a). The brightness gradually diminishes from the center outward, forming a typical fractal diffusion pattern, with highly intricate fractal details at the boundary.

When $\alpha_1=0.1+1.05i$, the geometric structure of the fractal system undergoes significant transformation in Fig.\ref{fig:1dSCS}(b). At this point, the fractal pattern exhibits strong rotational symmetry, with multiple spiral arms extending outward from the center, forming a complex spiral pattern. The appearance of the spiral structure reflects the system's nonlinear iterative characteristics under this parameter, while also revealing the high sensitivity of such fractals to parameter perturbations. The boundary still retains fractal self-similarity, but the pronounced rotational symmetry in the image indicates an enhancement in the symmetry and periodicity of the fractal.

When $\alpha_1=0.2+1.01i$, the fractal structure undergoes another transformation in Fig.\ref{fig:1dSCS}(c). The fractal becomes more compact, with the central luminous region becoming more concentrated, and the self-similarity recurs in a denser manner. Under this parameter, the recursion frequency and complexity of the fractal are significantly increased, presenting a dense multi-scale structure. Unlike when the complex parameter is $\alpha_1=0.1+1.01i$, the fractal system now displays a more compact distribution of details over a larger area. This suggests a notable impact of parameter adjustments on the geometric characteristics of the system, and a reduced sensitivity to small perturbations.

When $\alpha_1=0.5+1.05i$, the fractal system becomes more symmetrical, exhibiting an overall snowflake-like structure in Fig.\ref{fig:1dSCS}(d). This suggests that as the parameter continues to increase, the system tends towards greater symmetry and regularity, with a reduction in complexity. The detail density of the fractal remains high, but the luminous region at the center is more dispersed compared to previous fractal images, and the lengths of the fractal branches are more uniform. At this stage, the fractal exhibits a stable multi-scale recursion pattern, with a tendency for outward symmetrical expansion as the parameter increases.

\begin{figure}
	\centering
	\includegraphics[width=0.9\linewidth]{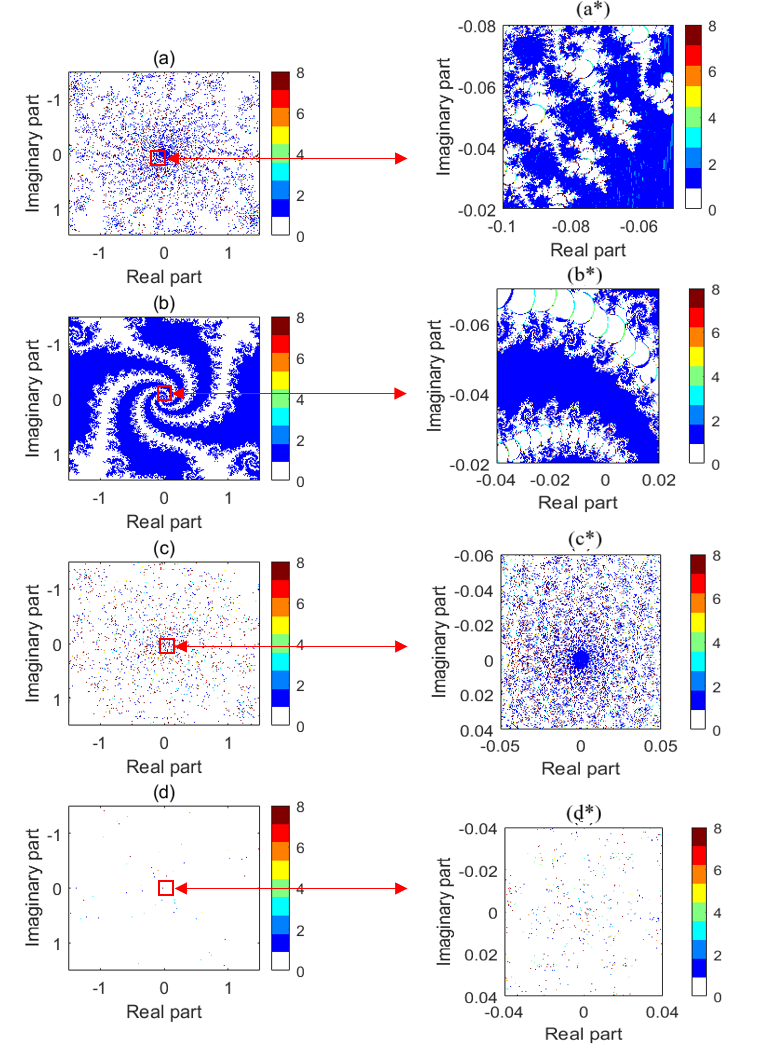}
	\caption{Julia Set with periodicity mapping of one-dimensional Sine-Cosine System
		(a)$\alpha_1=0.1+1.01i$
		(b)$\alpha_1=0.1+1.05i$
		(c)$\alpha_1=0.2+1.01i$
		(d)$\alpha_1=0.5+1.05i$}
	\label{fig:fencha}
\end{figure}

The periodicity mapping of the Julia set effectively characterizes the periodic behavior of a system under varying parameters. Periodicity distribution for one-dimensional SCDS with the four parameters are illustrated in Fig.\ref{fig:fencha}, where the figures on the left are the global view, and the figures on the right  are from an enlarged view and highlight the details of the periodic distribution within a selected region from the global map. The color coding reflects the length of the period, increasing from deep blue to red (with period-8 and above represented in dark red). White areas are not part of the Julia set.

In Fig.\ref{fig:fencha}(a) and Fig.\ref{fig:fencha}(b), a large connected region of stability is observed, predominantly with period-1 states, while other periodicities are mainly distributed along the boundaries of these connected regions. In Fig.\ref{fig:fencha}(c), the connected domain where the system is stable disappears. In Fig.\ref{fig:fencha}(d), the Julia set of the system is significantly reduced, with points of various periodicities sparsely distributed across the image.

\begin{figure*}
	\centering
	\includegraphics[width=1\linewidth]{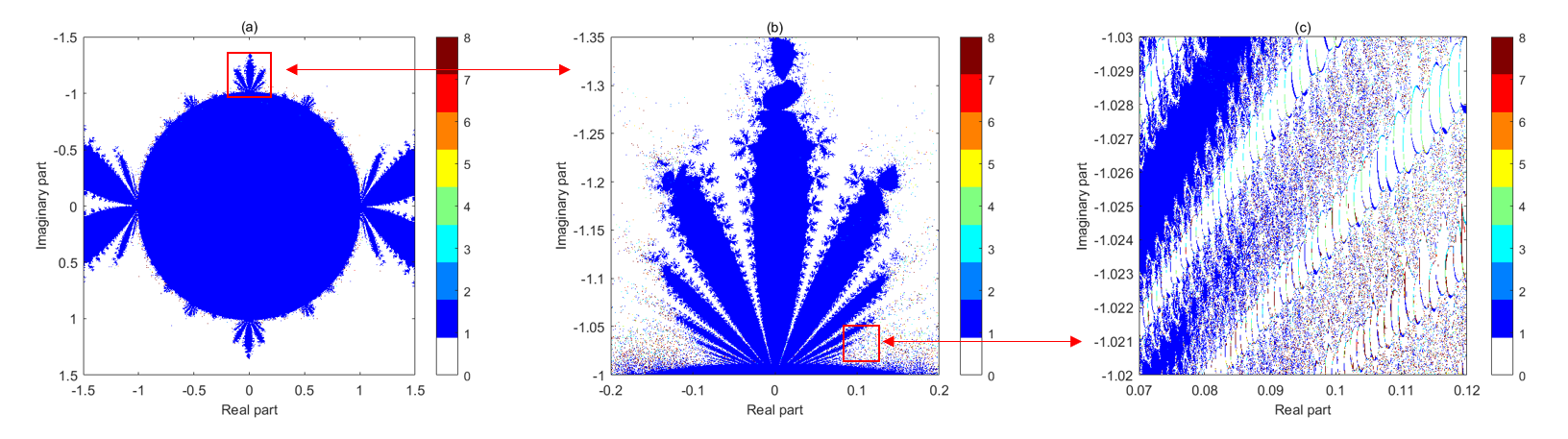}
	\caption{Bifurcation of one-dimensional Sine-Cosine System}
	\label{fig:tfencha1}
\end{figure*}

In Fig.\ref{fig:tfencha1}, the bifurcation diagram of the SCDS is shown, with initial condition $0.1 + 0.1i$. The system's behavior is analyzed by varying the parameter $\alpha_1$ in both the real and complex domains.

Fig.\ref{fig:tfencha1}(a) shows the bifurcation diagram of the entire system for different values of the parameter $\alpha_1$, where most regions exhibit a period-one state, indicating that the system behaves stably in a periodic manner within this range. The overall diagram displays relatively regular bifurcation features, showing orderly and smooth branches.

Fig.\ref{fig:tfencha1}(b) provides an enlarged view of a specific region from Fig.\ref{fig:tfencha1} (a). In this region, the bifurcation diagram takes on a shape reminiscent of a peacock's tail, with the boundaries between multiple periodic states displaying a more complex and symmetrical distribution with finer structures and local self-similarity. 

Fig.\ref{fig:tfencha1}(c) further zooms in on a part of Fig.\ref{fig:tfencha1} (b), showcasing the rich behavior of periodic states at the edges. In this localized region, the bifurcation diagram reveals numerous but disconnected periodic states, highlighting the system's high sensitivity and nonlinear characteristics under parameter variations. The alternating changes of these periodic states and their irregular distribution at the edges further demonstrate the chaotic nature and the complexity of the SCDS.

\begin{table}
\caption{\label{tab:table1} Fractal dimensions of one-dimensional SCDS under different parameters.}
\begin{ruledtabular}
	\begin{tabular}{lcccr}
		Parameter $\alpha_1$ & 0.1+1.01i & 0.1+1.05i & 0.2+1.01i & 0.5+1.05i\\
		\hline
		fractal dimension & 1.9800 & 1.8300 & 1.9879 & 1.9886\\
	\end{tabular}
\end{ruledtabular}
\label{table: m=1}
\end{table}

Fractal dimension quantifies the complexity and self-similarity of fractal objects, revealing how they fill space and their irregularity. We calculated the fractal dimension of the one-dimensional SCDS, as shown in Table \ref{table: m=1}. When $\alpha_1=0.1+1.01i$, the fractal dimension approaches 2, indicating that the fractal's geometric details are highly intricate, nearly filling the entire two-dimensional space, and the structure is very close to the complexity of a two-dimensional plane. Despite its intricate self-similarity, the fractal's details remain confined to the two-dimensional plane. The higher dimension reflects the fine structures that recur across multiple scales. When $\alpha_1=0.1+1.05i$, the fractal dimension drops to 1.8300, suggesting a reduction in the complexity of the fractal structure. However, when the real part increases to 0.2, the fractal dimension nearly returns to 2, indicating an increase in geometric complexity with different Julia sets.

Fig.\ref{fig:1dSCS}, Fig.\ref{fig:fencha}, Fig.\ref{fig:tfencha1} and Table \ref{table: m=1} exhibit a high degree of parameter sensitivity of one-dimensional SCDS. Small changes in parameters can lead to significant variations in the geometric structure and complexity of fractals, including the shape, size, the brightness, density, symmetry, self-similarity and so on.
Therefore, we can control parameters when working with fractals in various applications.

\subsection{\label{sec:level2} Chebyshev system in Complex Field}

\begin{figure}
\centering
\includegraphics[width=0.9\linewidth]{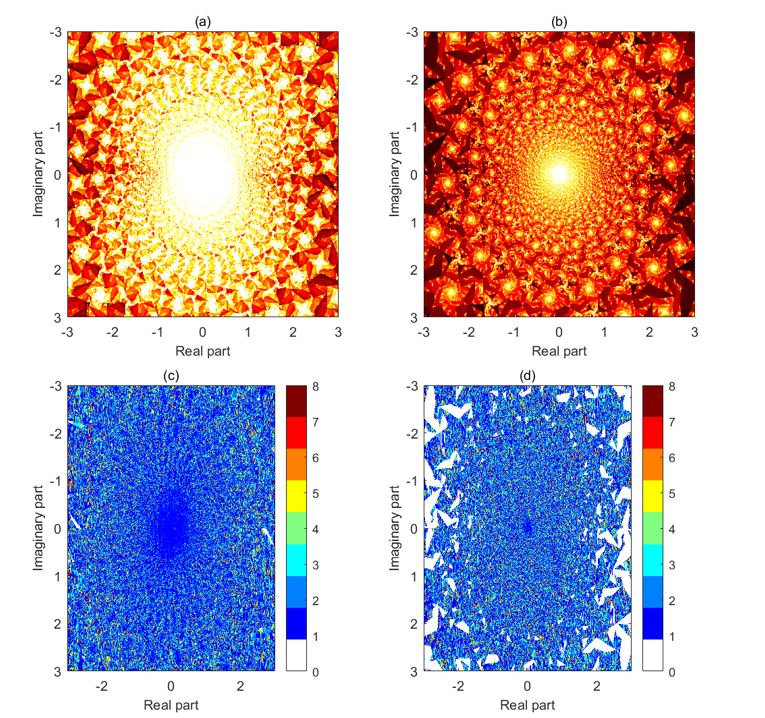}
\caption{Julia Set and periodicity mapping of the Julia Set of one-dimensional Chebyshev system,
	(a) and (c) $\alpha_1=1 + 0.15i$, 
	(b) and (d) $\alpha_1=1 + 0.2i$}
\label{fig:1dcs}
\end{figure}

Let $m$ = 1 and $\alpha_1$ take values of $1 + 0.15i$ and $1 + 0.2i$, respectively, yielding different Julia sets and Periodicity Mapping displayed in Fig.\ref{fig:1dcs}.

Overall, in Fig.\ref{fig:1dcs}(a)(c) and (b)(d), as the parameter $\alpha_1$ increases from $1+0.15i$ to $1+0.2i$, the fractal structure gradually evolves from a relatively loose central spiral pattern to a more compact and intricate petal-like distribution. Throughout this process, the system retains a complex periodic state. This transformation highlights the Chebyshev system's sensitivity to parameter variations, showcasing the intricate details and self-similar characteristics of its fractal geometry within the complex plane.

\begin{figure}
	\centering
	\includegraphics[width=1\linewidth]{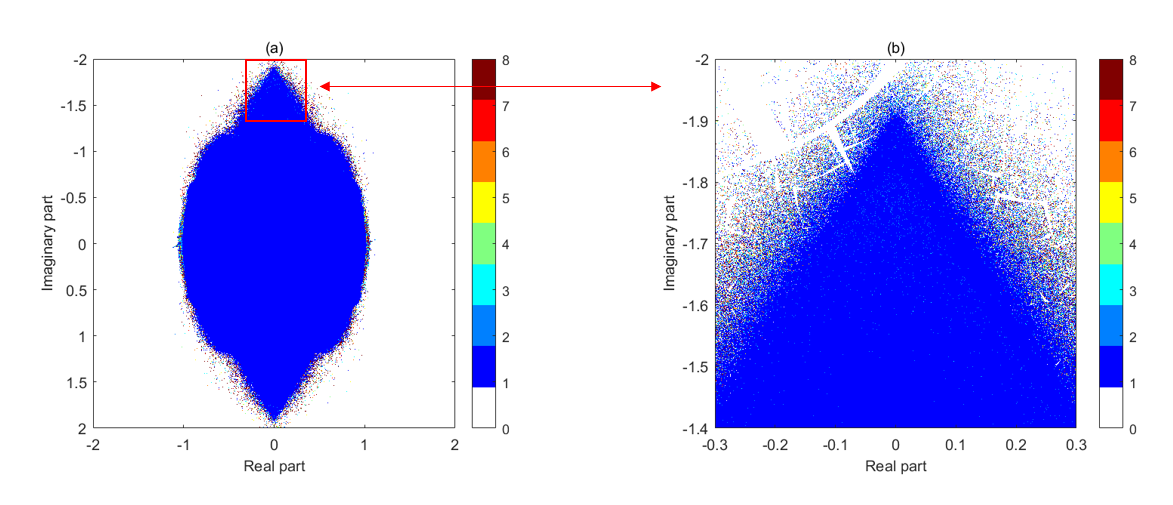}
	\caption{Bifurcation of one-dimensional Chebyshev system}
	\label{fig:tfencha2}
\end{figure}

Bifurcation of one-dimensional Chebyshev system is shown in Fig.\ref{fig:tfencha2}, exhibiting characteristics similar to those in Fig.\ref{fig:tfencha1}. As shown in Fig.\ref{fig:tfencha2}(a), the majority of the system's parameter space is dominated by period-one behavior, with a stable periodic state observed across most regions. In the zoomed-in Fig.\ref{fig:tfencha2}(b), a rich array of periodic states is displayed, though these states appear numerous yet disjointed, highlighting the system's sensitivity to parameter variations and the emergence of complex dynamics.

\subsection{\label{sec:level2} Sine-Logistic System in Complex Field}

\begin{figure}
	\centering
	\includegraphics[width=0.9\linewidth]{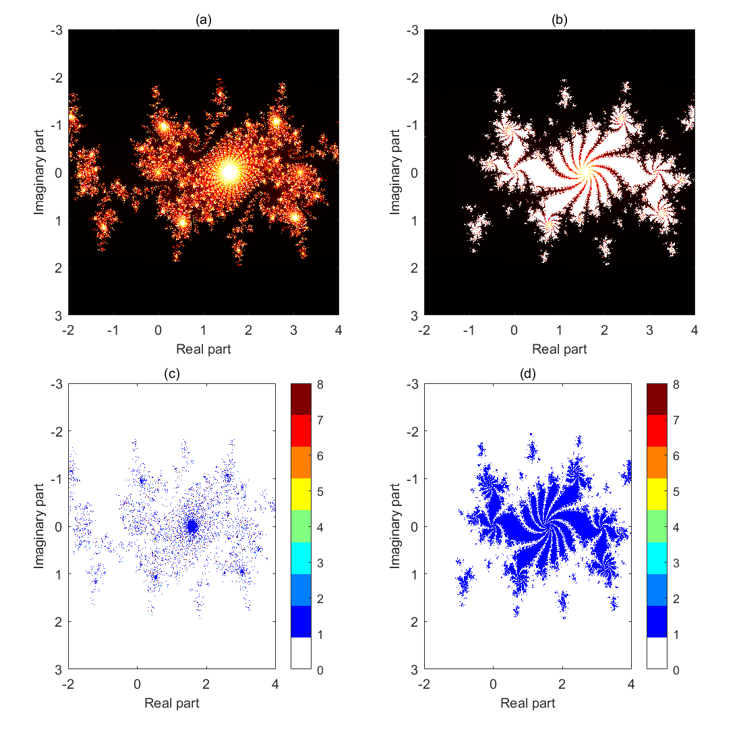}
	\caption{Julia Set and periodicity mapping of the Julia Set of Sine-Logistic System,
		(a) and (c) $\alpha_1=0.9 + 0.48i$, 
		(b) and (d) $\alpha_1=0.5 + 0.9i$}
	\label{fig:sls}
\end{figure}
\begin{figure*}
	\centering
	\includegraphics[width=1\linewidth]{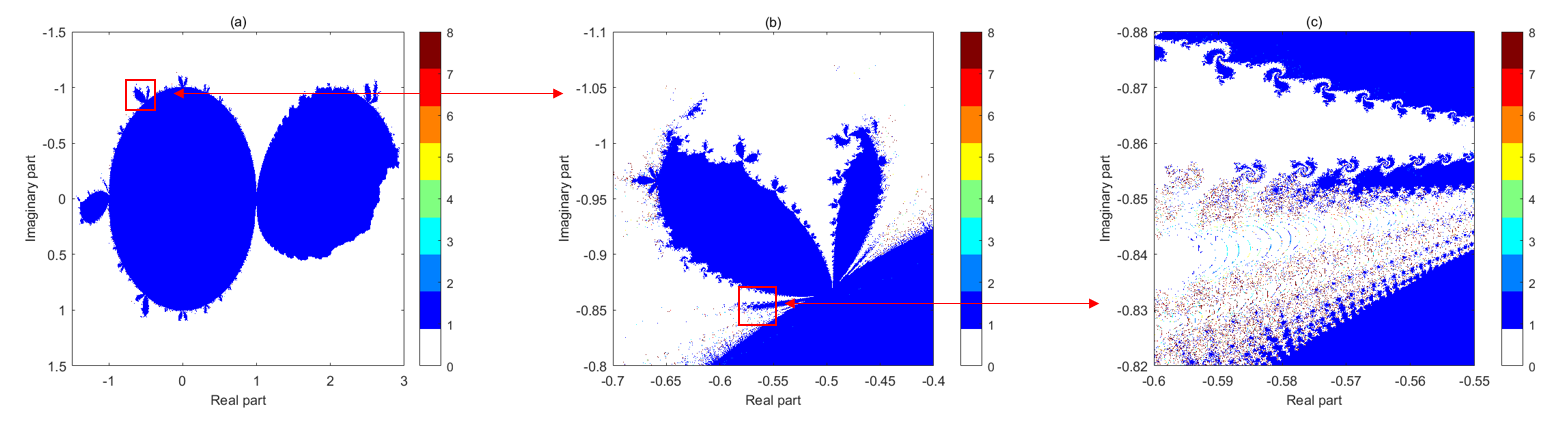}
	\caption{Bifurcation of Sine-Logistic System}
	\label{fig:tfencha3}
\end{figure*}

Similarly, for system (\ref{sls}), when $m = 1$, setting $\alpha_1$ as $0.9 + 0.48i$ and $0.5 + 0.9i$ yields the following results.

Bifurcation of Sine-Logistic System is shown in Fig.\ref{fig:sls}. These two images illustrate the fractal characteristics of the SLS with different $\alpha_1$, highlighting the system's sensitivity and diversity in response to parameter changes. When $\alpha_1 = 0.9 + 0.48i$, the fractal image exhibits a densely packed, nebula-like structure, with a bright, radially oriented spiral at the center that extends outward into intricate branching patterns. This structure reflects there exist central attraction and pronounced self-similarity in SLS. When $\alpha_1 = 0.5 + 0.9i$, the fractal image is characterized by a multi-armed, rotating star-like structure, with a central, spiraling symmetry that radiates outward into fine branches, which reveals the SLS's strong symmetry and rich dynamic extensibility. 

The bifurcation diagram of Sine-Logistic System is presented in Fig.\ref{fig:tfencha3}, exhibiting features similar to those in Fig.\ref{fig:tfencha1}. As in the previous case, the system's behavior is analyzed by varying the parameter $\alpha_1$ in both the real and complex domains, unveiling intricate and complex dynamics.

\subsection{\label{sec:level1}The generation mechanism of fractal for SCNSF}

The generation mechanisms of fractals for SCNSF are mainly twofold: iteration and bifurcation.

Fractals of SCNSF are generated through iterative processes. Beginning with an initial condition, a set of rules is repeatedly applied to produce a sequence of points or shapes. As the iterations continue, the resulting structure grows increasingly complex and exhibits self-similarity at different scales such as Fig.\ref{fig:1dSCS}, Fig.\ref{fig:fencha}, Fig.\ref{fig:1dcs}, Fig.\ref{fig:sls}, which is a distinctive characteristic of fractals. Take one-dimensional SCDS as an example, where $x_{n+1}=\alpha_1sin(x_n)$  is applied repeatedly. This iterative process is carried on for a certain number of times or until a desired pattern emerges. As the iteration progresses, the complexity and self-similarity of the resulting fractal become more evident. Small changes in the initial conditions or the mathematical rule can lead to significant differences in the final fractal structure. This sensitivity to initial conditions is a characteristic feature of fractals generated through iteration. The iterative generation of fractals can be visualized using computer programs or mathematical software. By observing the evolution of the fractal over successive iterations, one can gain a deeper understanding of the underlying mathematical principles and the beauty of fractal geometry.

On the other hand, bifurcation also plays a crucial role in the generation of fractals. As one system's parameters are varied, it can undergo bifurcations, where the system's behavior changes qualitatively. These bifurcations can lead to the emergence of complex and fractal structures such as Fig.\ref{fig:tfencha1}, Fig.\ref{fig:tfencha2}, Fig.\ref{fig:tfencha3}. 

In conclusion, both iteration and bifurcation are important mechanisms for the generation of fractals. Iteration provides a way to build complex structures through repeated application of simple rules, while bifurcation offers a route to the emergence of fractals through changes in the system's parameters. 

\section{\label{sec:level1} Conclusion}
In this paper, we propose Sine-Cosine Nonlinear System Family with a specific chaos mechanism in real field and fractal mechanism in complex field. The main conclusions are as follows: (1) We propose the Sine-Cosine Nonlinear System Family, which combines sine and cosine functions to generate chaos and fractals in both real and complex fields. (2) Based on the coupling effect of sine and cosine functions, three novel classes of sine-cosine nonlinear systems, including the Sine-Cosine Discrete System (SCDS), the Multidimensional Chebyshev System (MDCS), and the Sine-Logistic System (SLS), are proposed. They  exhibit complicated chaos and fractals characteristics with remarkable simplicity. This simplicity not only makes them conceptually accessible but also extremely easy for physical implementation, making them a promising choice for researchers and practitioners. (3) The mechanisms by which these systems generate chaos and fractal are proposed. They offer valuable insights into the underlying processes that lead to the emergence of chaos and fractals in these systems. Understanding these mechanisms can help us better understand the nature of chaos and fractals in various scientific and engineering fields.

In the future, we plan to conduct more in-depth and systematic research on spatial chaos and spatial fractals. Specifically, we will focus on exploring how chaotic and fractal structures are generated and evolve in higher-dimensional and more complex systems. This will include investigating multidimensional nonlinear systems, particularly their manifestations in physical and mathematical models. By introducing new computational methods and experimental techniques, we will uncover deeper connections between spatial chaos and fractal geometry, providing novel insights into the theoretical foundations of these complex phenomena. Additionally, we will analyze the potential applications of spatial chaos and spatial fractals, especially in areas such as information encryption, pattern recognition, and complex network dynamics. 

\begin{acknowledgments}
This work was supported by Shandong Province Natural Science Foundation (Nos. ZR2023MF089, ZR2023QF036 and ZR2022MF318) National Natural Science Foundation of China (Nos. 62172262 and 62406156) Shandong Province Small and Medium - sized Enterprise Innovation Capacity Improvement Project (2023TSGC039, 2024TSGC0645 and 2023TSGC0201) and the Program of Basic research projects of Qilu University of Technology(Shandong Academy of Sciences)(No.2023PX002 and No.2023PX081).
\end{acknowledgments}

\nocite{*}
\bibliography{ref}

\end{document}